\documentclass{article}
\usepackage{amssymb}

\usepackage{latexsym}
\usepackage{amsmath}
\usepackage{graphicx}
\usepackage{float}

\newlength{\dinwidth}
\newlength{\dinmargin}
\begin{document}

%--------------------------------------------------------------------------

%--------------------------------------------------------------------------
%--------------------------------------------------------------------------

\thispagestyle{empty} \vspace*{1cm}

\vspace*{2cm}

\begin{center}
{\LARGE A general CFT model for antiferromagnetic spin-1/2 ladders with
Mobius boundary conditions }

{\LARGE \ }

{\large Gerardo Cristofano\footnote{{\large {\footnotesize Dipartimento di
Scienze Fisiche,}\textit{\ {\footnotesize Universit\'{a} di Napoli
``Federico II''\ \newline
and INFN, Sezione di Napoli},}{\small Via Cintia, Compl.\ universitario M.
Sant'Angelo, 80126 Napoli, Italy}}}, Vincenzo Marotta\footnote{{\large
{\footnotesize Dipartimento di Scienze Fisiche,}\textit{\ {\footnotesize %
Universit\'{a} di Napoli ``Federico II''\ \newline
and INFN, Sezione di Napoli},}{\small Via Cintia, Compl.\ universitario M.
Sant'Angelo, 80126 Napoli, Italy}}},} {\large Adele Naddeo\footnote{{\large
{\footnotesize CNISM, Unit\`{a} di Ricerca di Salerno and Dipartimento di
Fisica \textit{''}E. R. Caianiello'',}\textit{\ {\footnotesize %
Universit\'{a} degli Studi di Salerno, }}{\small Via Salvador Allende, 84081
Baronissi (SA), Italy}}} \footnote{{\large {\footnotesize Dipartimento di
Scienze Fisiche,}\textit{\ {\footnotesize Universit\'{a} di Napoli
``Federico II''}, }{\small Via Cintia, Compl.\ universitario M. Sant'Angelo,
80126 Napoli, Italy}}}, Giuliano Niccoli\footnote{{\large {\footnotesize %
Theoretical Physics Group, DESY, NotkeStra\ss e 85 22603, Hamburg, Germany.}}%
}}

{\small \ }

\textbf{Abstract\\[0pt]
}
\end{center}

\begin{quotation}
We show how the low-energy properties of the 2-leg $XXZ$ spin-1/2 ladders
with general anisotropy parameter $\Delta $ on closed geometries can be
accounted for in the framework of the $m$-reduction procedure developed in
\cite{cgm4}. In the limit of quasi-decoupled chains, a conformal field
theory (CFT) with central charge $c=2$ is derived and its ability to
describe the model with different boundary conditions is shown. Special
emphasis is given to the Mobius boundary conditions which generate a
topological defect corresponding to non trivial single-spinon excitations.
Then, in the case of the 2-leg $XXX$ ladders we discuss\ in detail the role
of various perturbations in determining the renormalization group flow
starting from the ultraviolet (UV) critical point with $c=2$.

\vspace*{0.5cm}


{\footnotesize PACS: 11.25.Hf, 02.20.Sv, 03.65.Fd\newpage }\baselineskip%
=18pt \setcounter{page}{2}
\end{quotation}

\section{Introduction}

Since two decades one-dimensional and quasi one-dimensional
quantum spin systems have been the subject of an extensive study.
Such systems appear to be particularly interesting because of two
main features: Haldane's conjecture for one-dimensional
antiferromagnetic (AF) spin systems \cite {haldane} and the
discovery of ladder materials \cite{ladder}. In spin ladders two
or more spin chains interact with each other. The magnetic phases
of ladder systems are rich and strongly dependent on their
geometry, in particular several disordered ``quantum spin
liquid''\ phases are known \cite{liquid}. Due to the low
dimensionality quantum fluctuations are crucial and, because of
that, such systems exhibit a variety of interesting phenomena such
as the appearance of plateaux in magnetization curves \cite
{plateaux} and chiral spin liquid (CSL) phases corresponding to a
new class of non-Fermi-liquid fixed points \cite{csl}.

The study of spin ladders with different topologies of couplings has been
motivated both from the experimental and the theoretical side. In general,
ladders with even number of legs are disordered spin liquids with a finite
gap in the excitation spectrum, while in odd-legged ladders there exists a
gapless branch in the spectrum implying that the spin correlations decay
algebraically. The 3-leg frustrated ladder is the simplest model displaying
a non trivial CSL physics at the Toulouse point \cite{csl}. Relevant
examples of two-leg ladders are the railroad ladder and the zigzag ladder.

The railroad ladder \cite{railroad} has received a lot of attention in the
last years in relation to the physics of high $T_{c}$ cuprate
superconductors \cite{scalapino}; it can be thought of as a strip of a
two-dimensional square lattice. Within such a system it is possible to
observe how spin-1/2 excitations (spinons) are confined into magnons by
measuring the dynamical susceptibility $\chi ^{^{\prime \prime }}\left(
\omega ,q\right) $; the interchain exchange $J_{\perp }$ acts as a control
parameter and at $\left| J_{\perp }\right| \ll J_{||}$, where $J_{||}$ is
the intrachain coupling, there is a wide energy range where $\chi ^{^{\prime
\prime }}$ is dominated by incoherent multiparticle processes and a narrow
region at low energies where $\chi ^{^{\prime \prime }}$ exhibits a
single-magnon peak around $q=\pi $. So, a weakly coupled railroad ladder
gives the interesting opportunity to study the formation of massive spin $%
S=1 $ (triplet) and $S=0$ (singlet) particles, with gaps $m_{t}$ and $m_{s}$
($m_{t},m_{s}\sim J_{\perp }$), which appear as bound states of the spin-1/2
spinons of individual Heisenberg chains \cite{shelton1}. In this context it
has been shown that a quite strong four-spin interaction can induce
dimerization \cite{wang1}. The dimerized phases are thermodynamically
indistinguishable from the Haldane phase but the correlation functions show
a different behavior: the novel feature in the dynamical susceptibility is
the absence of a sharp single magnon peak near $q=\pi $, replaced by a
two-particle threshold separated from the ground state by a gap.

Much more interesting are the zigzag ladders which can be viewed as
one-dimensional strips of the triangular lattice and, then, could be used as
a toy model for understanding spin systems on such a lattice. Such ladders,
according to the value of the ratio between the leg and the rung exchange
couplings, may develop gapless ground states with algebraically decaying
spin correlations or spontaneously broken dimerized ground states \cite
{haldane1}. The gapped dimer ground state is degenerate with pairs of
spinons as elementary excitations \textbf{\cite{shastry} }. In the case of
antiferromagnetic (AF) couplings, the zigzag ladder introduces frustration,
which is a crucial and interesting feature, as shown in the following. Spin
ladders have been experimentally found in a number of quasi-one dimensional
compounds such as $SrCu_{2}O_{3}$ and $CuGeO_{3}$ \cite{exper}, $KCuCl_{3}$
and $TlCuCl_{3}$ \cite{exper1,exper2}. Zigzag coupling is also interesting
because, as shown later, it cancels the most relevant coupling term for the
usual ladders and the net result is a chirally asymmetric perturbation \cite
{chiral}. The study of such systems is relevant also in other contexts, such
as gated Josephson junction arrays \cite{gjja}. Furthermore, there should be
pointed out the close analogy \cite{white1} between the 2-leg zigzag ladder
and the doped Kondo lattice model, which is believed to be relevant to the
phenomenon of colossal magnetoresistance \cite{moxides} in metallic oxides.

The 2-leg spin-1/2 zigzag ladder is the simplest example of frustrated spin
systems and may help to elucidate the role of frustration and its interplay
with quantum fluctuations, in particular how it affects the spectrum of the
low energy excitations \cite{shastry}. Frustration is expected to lead to
new exotic phases as well as to unconventional spin excitations. Indeed such
a system can be reinterpreted as a $XXX$ Heisenberg chain with
next-to-nearest neighbor interaction, so the simplest Hamiltonian which
describes it is the isotropic one \cite{haldane1}:
\begin{equation}
H=\sum_{j}\left[ J_{1}\overrightarrow{S}_{j}\cdot \overrightarrow{S}%
_{j+1}+J_{2}\overrightarrow{S}_{j}\cdot \overrightarrow{S}_{j+2}\right] ,
\label{hais}
\end{equation}
where $\overrightarrow{S}_{j}\equiv \left( S_{j}^{\pm }=S_{j}^{x}\pm
iS_{j}^{y},S_{j}^{z}\right) $ is the spin operator at site $j$ and the
nearest and next-to-nearest neighbor exchanges $J_{1}$, $J_{2}>0$ are
competing antiferromagnetic interactions; in particular $J_{2}$ introduces
frustration in the model. Small values of $J_{2}$ give rise to a spin-fluid
phase whose effective theory is that of a massless free boson; it displays
quasi-long range spin order with algebraic decay of spin correlations.
Larger values of $J_{2}$ give rise to a quantum phase transition of
Kosterlitz-Thouless (KT) type characterized by a spontaneously dimerized
ground state.

Many differences arise in the strongly anisotropic $XX$ ladder case
described in terms of the Hamiltonian:
\begin{equation}
H_{XX}=\sum_{j}\left[ J_{1}\left(
S_{j}^{x}S_{j+1}^{x}+S_{j}^{y}S_{j+1}^{y}\right) +J_{2}\left(
S_{j}^{x}S_{j+2}^{x}+S_{j}^{y}S_{j+2}^{y}\right) \right] .  \label{haan}
\end{equation}
In such a case, in addition to the two previous phases, a new phase with
unconventional characteristics has been predicted in the large $J_{2}$ limit
\cite{chiral}, in particular a long-range chiral order
\begin{equation}
\left\langle \left( \overrightarrow{S}_{n}\wedge \overrightarrow{S}%
_{n+1}\right) _{z}\right\rangle \neq 0
\end{equation}
arises with local spin currents polarized along the anisotropy $z$ axis.
Such a phase with unbroken time reversal symmetry, termed spin nematic, is
gapless and displays incommensurate spin correlations which decay
algebraically with exponent $\frac{1}{4}$. In the ground state the
longitudinal and transverse spin currents are equal in magnitude but
propagate in opposite direction; that produces local currents circulating
around the triangular plaquettes of the ladder in an alternating way, the
total spin current of the system being zero.

Furthermore, in analyzing the low energy excitations of the zigzag ladder,
it is very interesting to investigate the nature of the single-spinon
excitation which may appear in the presence of a ``local topological
defect''\ in the ground state. Such an issue can be accounted for by
imposing Mobius boundary conditions (MBC) inside the closed ladder \cite
{mobius}; that creates a domain-wall type defect in the system without
loosing translational invariance. The topological defect can move around in
the system and gives rise to single-spinon excitations. That is, non trivial
boundary conditions induce interesting new properties in the behavior of the
ladder system modifying its excitation spectrum: different boundary
conditions on the spin chains give rise to a different spectrum of
excitations both within theoretical and numerical analysis \cite{grimm}. The
implications of closed geometries appear to be very exciting, also in view
of the study of new kinds of topological order and fractionalized phases
\cite{topological}\cite{wen}, not yet well explored in the existing
literature.

The aim of this paper is to analyze the ground states and the low-energy
excitations of a class of antiferromagnetic 2-leg spin-1/2 $XXZ$ Heisenberg
ladders arranged in a closed geometry for a variety of boundary conditions,
employing a twisted CFT approach, the twisted model (TM) \cite{cgm4}. Such
an approach has been successfully applied to quantum Hall systems in the
presence of impurities or defects \cite{noi1,noi2,noi5}, to Josephson
junction ladders and arrays of non trivial geometry, in order to investigate
the existence of topological order and magnetic flux fractionalization in
view of the implementation of a possible solid state qubit protected from
decoherence \cite{noi3,noi4,noi6} and, finally, to the study of the phase
diagram of the fully frustrated $XY$ model ($FFXY$) on a square lattice \cite
{noi}. In this way we build up a complete CFT for the closed 2-leg $XXZ$
spin-1/2 ladders with general anisotropy parameter $\Delta $, which captures
the universal features, reproduces the basic phenomenology and also embodies
non trivial boundary conditions. In such a context we focus on the weak
coupling limit between the two legs of the ladder, well described by a CFT
with central charge $c=2$. Then we restrict to the $XXX$ case and analyze
the role of various perturbations in determining the renormalization group
flow to several infrared (IR) fixed points in order to reproduce all the
relevant phenomenology.

The paper is organized as follows. In Section 2, we introduce the
lattice models under study; we start with two non interacting
$XXZ$ spin-1/2 chains, characterized by a general value of the
anisotropy parameter $\Delta $ and concentrate our attention on
the topological conditions which arise from a closed geometry. We
focus on two kinds of boundary conditions, that is periodic
boundary conditions (PBC) and Mobius boundary conditions (MBC) and
analyze its implications on the spectrum and the low energy
excitations of the two chains system. Then we switch to the
interacting system by turning on the ladder perturbations; we
introduce four different perturbations, the railroad one, the
zigzag, the 4-spin and, finally, the 4-dimer one and briefly
recall all the relevant phenomenology with an emphasis on the two
particular values of the anisotropy parameter $\Delta =1$ (i.e.
the $XXX$ ladder) and $\Delta =0$ (i.e. the $XX$ ladder)
respectively. The implications of a non trivial geometry are also
discussed in this weakly interacting case. In Section 3, we recall
the continuum limit of the single antiferromagnetic $XXZ$ spin-1/2
chain making use of an Abelian bosonization \cite{bos} framework.
We focus on the exact relation, given by Bethe Ansatz
\cite{betheansatz}, between the compactification radius $R_{\phi
}$ of the boson theory and the anisotropy parameter $\Delta $ of
the chain which will be crucial in determining the symmetry
properties of TM. In Section 4, we recall some aspects of the
$m$-reduction procedure \cite{VM}, focusing in particular on the
$m=2$ case which is the one relevant for the 2-leg system. For the
first time we explicitly develop such a procedure for the scalar
case and evidence its peculiarities for a general compactification
radius. In Section 5, we show how the TM, generated by $m=2$%
-reduction, well describes in the continuum limit the 2-leg $XXZ$
spin-1/2 ladder with general anisotropy parameter $\Delta $
arranged in a closed geometry for periodic (PBC) and Mobius
boundary conditions (MBC). Then the symmetry properties of the TM
are recalled for the particular value of the compactification
radius corresponding to the $XXX$ ladder. In Section 6, we
introduce the interacting system in the continuum by turning on
four different perturbations, the railroad one, the zigzag, the
4-spin and the 4-dimer one. We restrict our analysis to the $XXX$
ladder and study the different renormalization group (RG)
trajectories obtained starting from the UV fixed point, described
by a CFT with central charge $c=2$. Depending on the perturbing
term we obtain different infrared (IR) fixed points corresponding
to different physical behaviors. In Section 7, some comments and
outlooks are given. In Appendix A, the low energy excitations on
the torus topology are explicitly given for the TM describing the
$XXX$ ladder. In Appendix B, we give a derivation of Eq.
(\ref{per29}) quoted in Section 6 within a self-consistent mean
field approximation.

\section{2-leg $XXZ$ spin-1/2 ladders with Mobius boundary conditions: the
lattice model}

In this Section we introduce in whole generality the system under study,
that is two antiferromagnetic spin-1/2 $XXZ$ chains arranged in a closed
geometry, and discuss the different topological conditions which arise
corresponding to different boundary conditions imposed at the ends of the
two chains. Then we switch to the interacting case and describe the
following perturbations: railroad, zigzag, 4-spin and 4-dimer. We discuss in
detail the two cases: $XXX$ ladder and $XX$ ladder.

The starting point is a system of two non interacting spin-1/2 $XXZ$ chains
with Hamiltonian:
\begin{equation}
H_{0}=J_{0}\sum_{i=1}^{M}(S_{i}^{x}S_{i+2}^{x}+S_{i}^{y}S_{i+2}^{y}+\Delta
S_{i}^{z}S_{i+2}^{z})  \label{h1}
\end{equation}
where $J_{0}>0$ is the antiferromagnetic coupling, $\Delta $ is the
anisotropy parameter, $M$ is the total number of sites and we assume that
the odd sites belong to the down leg while the even sites belong to the up
leg, as sketched in Fig. 1.
\begin{figure}[tbp]
\centering\includegraphics*[width=0.5\linewidth]{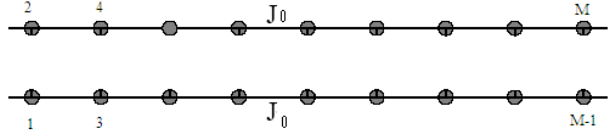}
\caption{The non interacting system}
\label{figura1}
\end{figure}
At this stage the two legs are not interacting. Let us now close the chains
imposing the following boundary conditions:
\begin{equation}
M+1\equiv 1\text{ \ , \ \ }M+2\equiv 2.  \label{bc1}
\end{equation}
Depending on $M$, we get two topologically inequivalent boundary conditions:
1) for $M$ even, we get two independent $XXZ$ spin-1/2 chains, each one
closed by gluing opposite ends: periodic boundary conditions (PBC); 2) for $%
M $ odd the two legs are not independent, indeed they appear to be connected
at a point upon gluing the opposite ends and the system can be viewed as a
single eight shaped chain. This local coupling is a source of local
incommensuration, even in the absence of further interactions between up and
down legs. Indeed, around the gluing point, an angle deviation is induced
between the spins on the two chains from the decoupled value $\theta =\frac{%
\pi }{2}$ \cite{white1}. That defines a characteristic wavelength which is
expected to be proportional to a finite correlation length $\xi _{spiral}$, $%
\theta -\frac{\pi }{2}\propto 1/\xi _{spiral}$, so it could give rise
locally to a finite range spiral order. That is, for $M$ odd the system
presents some kind of local topological defect in the gluing point,
described by:
\begin{equation*}
H_{0}^{MBC}=H_{0}+H_{MBC}
\end{equation*}
where the localized crossed interaction $H_{MBC}$ around the gluing point is
explicitly given by:
\begin{equation}
H_{MBC}=J_{0}\sum_{i=M,M+1}(S_{i}^{x}S_{i+1}^{x}+S_{i}^{y}S_{i+1}^{y}+\Delta
S_{i}^{z}S_{i+1}^{z})-J_{0}(S_{M}^{x}S_{M+2}^{x}+S_{M}^{y}S_{M+2}^{y}+\Delta
S_{M}^{z}S_{M+2}^{z}).  \label{mobiusbc1}
\end{equation}
This local deformation realizes the so called\ Mobius boundary conditions
(MBC) in the system by physically introducing a topological defect at the
gluing point. The system now exactly coincides with two closed $XXZ$
spin-1/2 chains, each one with $(M+1)/2$ sites, which intersect each other
in the topological defect at site $M+1\equiv 1$. The existence of two
topologically inequivalent configurations, which in the continuum are
described by the two topological sectors, untwisted and twisted one, of our
TM theory with central charge $c=2$ (see Section 5), is crucial in order to
recognize an underlying topological order \cite{noi7}.

Now we are ready to introduce interactions between the two spin-1/2 $XXZ$
chains. Let us start from those which involve two spins, the railroad
interaction:
\begin{equation}
H_{\text{Railroad}}=J_{\perp
}^{R}\sum_{i=1}^{[(M+1)/2]}(S_{2i-1}^{x}S_{2i}^{x}+S_{2i-1}^{y}S_{2i}^{y}+%
\Delta S_{2i-1}^{z}S_{2i}^{z}),  \label{p1}
\end{equation}
and the zigzag one:
\begin{equation}
H_{\text{ZigZag}}=J_{\perp
}^{Z}\sum_{i=1}^{M}(S_{i}^{x}S_{i+1}^{x}+S_{i}^{y}S_{i+1}^{y}+\Delta
S_{i}^{z}S_{i+1}^{z}).  \label{p2}
\end{equation}
In order to briefly recall the relevant phenomenology due to such
interactions, let us start with $H_{0}+H_{\text{Railroad}}$ \cite{railroad}
and discuss the isotropic $\Delta =1$ case (the system under study is
depicted in Fig. 2).

\begin{figure}[tbp]
\centering\includegraphics*[width=0.5\linewidth]{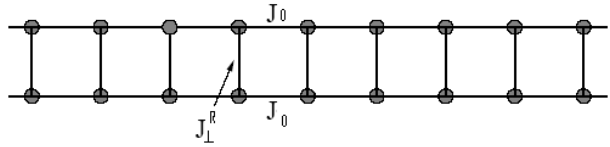}
\caption{The railroad ladder}
\label{figura2}
\end{figure}
The weak coupling limit $J_{0}\gg \left| J_{\perp }^{R}\right| $ is
interesting because it allows to study the crossover regime between the
gapless spinon excitations on the two decoupled spin-1/2 chains and the
strong coupling limit, which takes place when the energy scale is lowered
\cite{shelton1}. The main universal features of the spectrum appear to be
its symmetry and the persistence of a gap. Such a regime gives the
opportunity to investigate the formation of massive spin $S=1$ and $S=0$
particles, which appear as bound states of the spin-1/2 excitations of the
individual chains. Indeed, at small interchain coupling $\left| J_{\perp
}^{R}\right| \ll J_{0}$ the masses of such particles are of the order of $%
\left| J_{\perp }^{R}\right| $, the $S=1$ branch being always
lower. In the limit $\frac{J_{\perp }^{R}}{J_{0}}\rightarrow 0$
the singlet spectral gap results three times as large as the
triplet one. Conversely, in the opposite limit $J_{\perp }^{R}\gg
J_{0}$ the ground state is a product of rung singlets with a gap
to the excited states which can be viewed as the energy needed in
order to break a singlet bond. For $J_{\perp }^{R}\simeq J_{0}$
the energy gap still exists and the ground state is described by a
short range valence-bond state \cite{shelton1}\cite{nomura}. In
the strongly anisotropic case $\Delta =0$, for $J_{\perp }^{R}\gg
J_{0}$ we get a rung singlet massive phase with a non degenerate
ground state which is the direct product of singlets in the rung
direction. Such a phase extends in the whole
range of $J_{\perp }^{R}$, from very large values to very small ones, $%
J_{\perp }^{R}\simeq 0$ \cite{nomura}.

The zigzag ladder\ with Hamiltonian $H_{0}+H_{\text{ZigZag}}$ \cite{white1}
\cite{allen}\cite{chiral} \cite{itoi} (the system is shown in Fig. 3) is
equivalent to the spin-1/2 frustrated Heisenberg chain with nearest neighbor
coupling $J_{\perp }^{Z}$ and next to nearest neighbor coupling $J_{0}$.
\begin{figure}[tbp]
\centering\includegraphics*[width=0.5\linewidth]{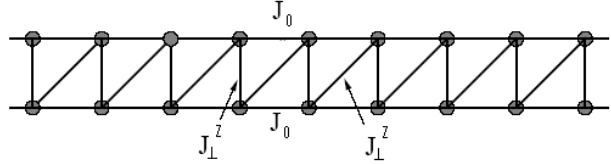}
\caption{The zigzag ladder}
\label{figura3}
\end{figure}
In the isotropic case $\Delta =1$, such a model is shown to be critical for $%
J_{\perp }^{Z}>J_{\perp c}^{Z}$ ($J_{\perp c}^{Z}=J_{0}/0.241$) \cite
{zigcritical}, being in the same universality class as the antiferromagnetic
spin-1/2 Heisenberg chain. For $J_{\perp }^{Z}<J_{\perp c}^{Z}$ the spinons
acquire a gap but remain still deconfined and the ground state becomes
doubly degenerate \cite{haldane1}. Finally, at the Majumdar-Ghosh (MG) point
\cite{ghosh} $J_{\perp }^{Z}=2J_{0}$, that is the exact dimerization point,
the two degenerate ground states assume a simple form in the thermodynamic
limit, being a collection of decoupled singlets as sketched in Fig. 4.

\begin{figure}[tbp]
\centering\includegraphics*[width=0.5\linewidth]{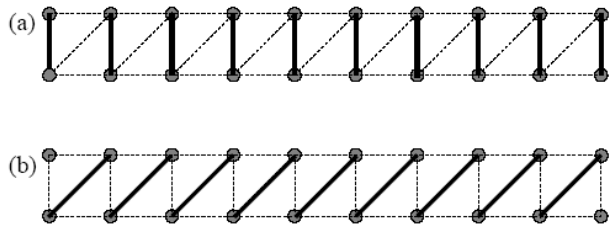}
\caption{The two degenerate ground states of the zigzag ladder at the MG
point}
\label{figure4}
\end{figure}
Such ground states can be continuously related to the rung singlet and
Haldane phases \cite{kolezuk}. For weaker interchain couplings $J_{\perp
}^{Z}<2J_{0}$ incommensurate spiral correlations appear in the short range
correlations \cite{white1}\cite{incommensurate}. In the regime $J_{0}\gg
J_{\perp }^{Z}$ we may view the system as two spin chains with a weak zigzag
interchain coupling \cite{white1}\cite{allen}: an exponentially small gap
develops and the weak interchain correlations break translational symmetry.
So this phase shows a spontaneous dimerization along with a finite range
incommensurate magnetic order. Exact diagonalizations as well as density
matrix renormalization group (DMRG) calculations \cite{sorensen} have
evidenced the difference between a zigzag ladder with an even and an odd
number $M$ of sites respectively in the regime $J_{0}\gg J_{\perp }^{Z}$. In
particular, for $M$ odd the $S=\frac{1}{2}$ ground state has a dispersion
relation with a well defined single particle mode around $k=\frac{\pi }{2}$:
such a particle can be viewed as a soliton whose gap increases exponentially
when $J_{0}>$ $0.241$. Furthermore no soliton-antisoliton bound states form
in this model and solitons are repelled by non magnetic impurities, that is
by the ends of the open chains. Conversely, for $M$ even at the MG point
there are exact singlet and triplet bound states in a small range of momenta
close to $q=\frac{\pi }{2}$, which are degenerate.

In the strongly anisotropic case $\Delta =0$, two coupled $XX$ chains \cite
{chiral}, the phase diagram is quite rich. Indeed for a small value of $%
J_{0} $ a spin fluid phase develops, characterized by gapless excitations
with central charge $c=1$, while for $\frac{J_{0}}{J_{\perp }^{Z}}\simeq
0.32 $ a phase transition of Kosterlitz-Thouless (KT) kind takes place and
the system enters a massive dimerized phase with a twofold degenerate ground
state. Conversely, in the large $J_{0}$ limit a critical spin nematic phase
with chiral long range order develops. Such a non trivial phase with
unbroken time-reversal symmetry is characterized by nonzero local spin
currents polarized along the anisotropy $z$ axis. It preserves the spin $%
U\left( 1\right) $ symmetry but spontaneously breaks the $Z_{2}$ symmetry.
The transverse spin-spin correlation functions are incommensurate and fall
off with the distance as a power law with exponent $\frac{1}{4}$. The ground
state is characterized by longitudinal and transverse spin currents which
are equal in magnitude but propagate in opposite directions. As a result
local currents circulating around the triangular plaquettes of the ladder in
an alternating way develop, the total spin current of the system being zero.
That gives rise to two degenerate ground states characterized by an
antiferromagnetic pattern of chiralities in adjacent plaquettes of the
ladder, as shown in Fig. 5.

\begin{figure}[tbp]
\centering\includegraphics*[width=0.5\linewidth]{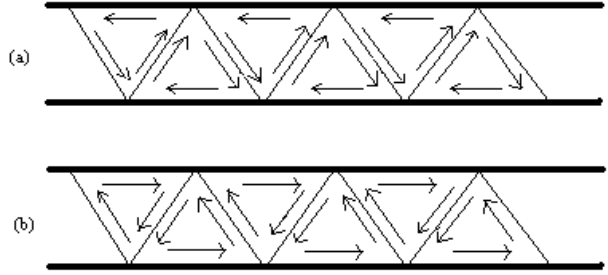}
\caption{The two degenerate ground states of the zigzag $XX$ ladder in the
spin nematic phase}
\label{figure5}
\end{figure}
Let us now introduce the closed geometry and point out the peculiarities
which emerge in the interacting system for the two inequivalent
configurations, $M$ even and odd. For $M$ even (PBC), the up and down legs
of the ladder are still well distinguished even if coupled by the two
interactions $H_{\text{Railroad}}$ and $H_{\text{ZigZag}}$; in this way the
ground state structure and the nature of the excitations is preserved.
Conversely, very different and more interesting features take place for $M$
odd (MBC). Indeed the zigzag interaction $H_{\text{ZigZag}}$ couples the $M$
site on the down leg with the $M+1\equiv 1$ site which still belongs to the
down leg.

In the $\Delta =1$ case, that breaks the ground states structure
and creates a domain-wall type defect in the system without
loosing the translational invariance; such a topological defect
can be viewed as a $S=\frac{1}{2}$ particle which is able to move
with finite momentum. It has been shown \cite {mobius} by Lanczos
and Householder diagonalization methods that the defect particle
gives rise to a single-spinon excitation in the spectrum. The
presence of an isolated soliton in odd-length frustrated chains with a $%
S_{tot}=\frac{1}{2}$ ground state has been evidenced by DMRG
methods as well \cite{sorensen}; in such a case the soliton
excitation behaves as a massive free particle which is repelled by
the ends of the chain, so it appears much more similar to a
massive particle in a box. In the $\Delta =0$ case, we have that
the alternating chiralities structure of the two ground states
gets lost. Such a feature can be viewed as the localization of a
topological defect which carries an elementary quantum of current,
in full analogy with the half flux quantum associated with the
topological defect in the closed Josephson junction ladder with
MBC investigated in Refs. \cite {noi3,noi4,noi6}: in this way the
condition of zero total spin current in the ladder would be
recovered. In presence of the railroad interaction we obtain more
drastic effects, indeed the MBC turn locally on the $M+1\equiv 1$
site the interaction into a zigzag one. More precisely, the $1$
site interacts both with the $M$ site and the $2$ site and also
with the $M-1$ site and the $3$ site via the Hamiltonian $H_{0}$.

The two spin interactions just introduced, $H_{\text{Railroad}}$ and $H_{%
\text{Zig-Zag}}$, can be described by means of a suitable unbalance of the
zigzag interaction term. That is interesting in the continuum limit, because
it is always possible to put together $H_{\text{Railroad}}$ and $H_{\text{%
Zig-Zag}}$ in a way to drop the marginal term which appears in $H_{\text{%
Railroad}}$ and obtain a pure relevant perturbation with conformal
dimensions $(\frac{1}{2},\frac{1}{2})$ \cite{shelton1}.

Let us now introduce the 4-spin interaction \cite{wang1}:
\begin{equation}
H_{\text{4-Spin}}=J_{\epsilon }\sum_{i=1}^{[(M+1)/2]}\epsilon
_{i}^{Dw}\epsilon _{i}^{Up},  \label{p3}
\end{equation}
which represents an interchain coupling of the spin-dimerization fields
defined on the two legs of the ladder by:
\begin{equation}
\begin{array}{c}
\epsilon
_{i}^{Dw}=(-1)^{i}(S_{2i-1}^{x}S_{2i+1}^{x}+S_{2i-1}^{y}S_{2i+1}^{y}+\Delta
S_{2i-1}^{z}S_{2i+1}^{z}) \\
\epsilon
_{i}^{Up}=(-1)^{i}(S_{2i}^{x}S_{2i+2}^{x}+S_{2i}^{y}S_{2i+2}^{y}+\Delta
S_{2i}^{z}S_{2i+2}^{z})
\end{array}
.  \label{p3a}
\end{equation}
That can be generated either by phonons or (in the doped state) by the
conventional Coulomb repulsion between the holes.

Finally, let us introduce the following 8-spin interaction, termed 4-dimer:
\begin{equation}
H_{\text{4-Dimer}}=J_{\varkappa }\sum_{i=1}^{[(M+1)/2]}\varkappa
_{i}\varkappa _{i+1},\text{ \ \ \ where: }\varkappa _{i}=\epsilon
_{i}^{Dw}\epsilon _{i}^{Up}\text{.}  \label{p4}
\end{equation}
It plays a crucial role in the presence of the marginal interaction $H_{%
\text{Zig-Zag}}$ only, in that it gives rise to the dynamical generation of
the triplet or singlet mass respectively along two different RG flows, in
full analogy with the previous case of the relevant perturbations $H_{\text{%
Railroad}}$ and $H_{\text{4-Spin}}$.

The continuum limit of all interaction terms just introduced together with
the different RG flows which originate by perturbing the two chain
Hamiltonian $H_{0}$ (or $H_{0}^{MBC}$) is discussed in detail in Section 6,
the continuum limit of $H_{0}$ (or $H_{0}^{MBC}$) being the subject of
Section 5.

\section{The XXZ spin-1/2 chain in the continuum limit}

In this Section we briefly recall the Abelian bosonization technique for the
description of the $XXZ$ spin-1/2 chain in the continuum limit. This quantum
field theory is the starting point (\textit{mother theory}, see next
Section) in our TM description in the continuum of the 2-leg $XXZ$ spin-1/2
ladders arranged in a closed geometry. Of particular interest for us is the
exact relation \cite{betheansatz} between the compactification radius $R$ of
the boson theory and the anisotropy parameter $\Delta $ of the chain; indeed
the symmetry properties of TM crucially depend on this compactification
radius. There is a wide literature on the subject of bosonization and on its
applications, a review can be found in Ref. \cite{bos}; here, we summarize
for clarity sake the main steps in the derivation \cite{Giam}.

Let us consider the single $XXZ$ spin-1/2 chain without magnetic field:
\begin{equation}
H_{XXZ}=J\sum_{j=1}^{N}(S_{j}^{x}S_{j+1}^{x}+S_{j}^{y}S_{j+1}^{y}+\Delta
S_{j}^{z}S_{j+1}^{z}),  \label{ha1}
\end{equation}
where $j$ is the site index, $N$ is the number of sites, $J>0$ is the
exchange coupling and $\Delta $ is the anisotropy parameter. We can convert
the above Hamiltonian to one of spinless interacting fermions on a lattice
via the Jordan-Wigner transformation. Then, we linearize the energy spectrum
around the Fermi surface and switch to the continuum limit by introducing
right and left-moving bosons $\phi_{R}\left( x\right)$ and $\phi_{L}\left(
x\right)$ (where $x=a j$, for lattice spacing equal to $a $). In this way
the continuum limit of the Hamiltonian (\ref{ha1}) reads:
\begin{equation}
H_{LL}=\frac{v_{F}}{8\pi }\int dx\left\{ \left( 1+g\right) \left[ \left(
\partial _{x}\phi _{R}\right) ^{2}+\left( \partial _{x}\phi _{L}\right) ^{2}%
\right] -2g\partial _{x}\phi _{L}\partial _{x}\phi _{R}\right\} ,
\label{ha2bis}
\end{equation}
where only the marginal operators have been kept (see Section 3 of Ref. \cite
{A-M} for more details) and $g=\left( 2J\Delta \sin ^{2}k_{F}\right) /\pi
v_{F}$, $v_{F}$\ being the Fermi velocity. This is the exactly solvable
Luttinger Hamiltonian:
\begin{equation}
H_{LL}=\frac{v}{8\pi }\int dx\left[ \left( \partial _{x}\theta \right)
^{2}+\left( \partial _{x}\phi \right) ^{2}\right] ,  \label{ha3}
\end{equation}
having introduced the bosonic field $\phi $ and its dual $\theta $, defined
as:
\begin{equation}
\phi =\frac{\phi _{L}+\phi _{R}}{\sqrt{K}},\text{ }\theta =\sqrt{K}(\phi
_{L}-\phi _{R}).  \label{dualfield2}
\end{equation}
Here, the renormalized velocity $v$ and the Luttinger parameter $K$ are
simply related to the parameters $g$ and $v_{F}$\ by:
\begin{equation}
v=\frac{v_{F}}{2}\sqrt{1+2g},\text{ }K=1/\sqrt{1+2g}.  \label{luttingerpar}
\end{equation}
Moreover, they can be expressed in terms of the \textit{microscopic}
parameters $J$ and $\Delta $ by comparison with exact Bethe Ansatz
calculations \cite{betheansatz}:
\begin{equation}
v\left( \Delta \right) =\frac{J\pi }{2}\frac{\sqrt{1-\Delta ^{2}}}{\arccos
\Delta },\text{ \ \ \ \ \ }K\left( \Delta \right) =\frac{\pi }{2\left( \pi
-\arccos \Delta \right) }.  \label{luttingerpar1}
\end{equation}
In the standard bosonization framework, the spin operators have the
following bosonic counterpart:
\begin{equation}
S_{j}^{z}\sim \sqrt{K}\partial _{x}\phi +(-1)^{j}const\times \cos (\sqrt{K}%
\phi ),\text{ \ \ }S_{j}^{-}\sim e^{-\frac{i}{2\sqrt{K}}\theta }\left[
(-1)^{j}+\cos (\sqrt{K}\phi )\right] ,
\end{equation}
which, under the assumption of periodic boundary conditions, imply the
compactification of the boson fields:
\begin{eqnarray}
\phi (x+L,t) &=&\phi (x,t)+2\pi m_{\phi }R_{\phi },\text{ \ for }m_{\phi
}\in \mathbb{Z}\text{ and }R_{\phi }=1/\sqrt{K},  \label{Comp-phi} \\
&&  \notag \\
\theta (x+L,t) &=&\theta (x,t)+2\pi m_{\theta }R_{\theta },\text{ \ for }%
m_{\theta }\in \mathbb{Z}\text{ and }R_{\theta }=2\sqrt{K}.
\end{eqnarray}
The compactification radii are functions only of the anisotropy parameter $%
\Delta $ while the Luttinger velocity $v$ (and analogously the lattice
coupling $J$) enters as an overall factor in the spectrum. Such a
consideration plays a crucial role in Section 5, in the construction of our
TM and its relation with the $XXZ$ 2-leg ladder system. It is worth pointing
out that the same\footnote{%
The compactification radius in \cite{L-T} coincides with our $R_{\theta }$,
up to a factor $\sqrt{2}$\ due to the different normalization of the boson
fields.} compactification conditions can be derived on the basis of pure
symmetry considerations \cite{L-T}.

Let us complete this Section by giving the mode expansion of the
compactified boson field\footnote{%
Within our normalization, the exponential field $V_{\alpha }(z,\bar{z}%
)=:e^{i\alpha \phi (z,\bar{z})}:$ has conformal dimensions $(\alpha
^{2}/2,\alpha ^{2}/2)$.}:
\begin{equation}
\phi (z,\bar{z})=q-ia_{0}\ln z-i\bar{a}_{0}\ln \bar{z}+\sum_{n\neq 0}\frac{%
ia_{n}}{n}z^{-n}+\sum_{n\neq 0}\frac{i\bar{a}_{n}}{n}\bar{z}^{-n}
\label{Boson-phi}
\end{equation}
in terms of the complex coordinates:
\begin{equation}
z=e^{2\pi (\tau -ix)/L},\text{ \ \ \ \ \ }\bar{z}=e^{2\pi (\tau +ix)/L},
\end{equation}
where $\tau =ivt$ (Euclidean space-time). Here, the zero modes are:
\begin{equation}
a_{0}=p+w\text{ and }\bar{a}_{0}=p-w
\end{equation}
where $w$ is the winding operator with eigenvalues $m_{\phi }R_{\phi }/2$
for $m_{\phi }\in \mathbb{Z}$ ($m_{\phi }$ is the \textit{winding number)}, $%
p$ is the conjugate momentum of $q$, $\left[ q,p\right] =i,$ and the $a$%
-modes satisfy two independent Heisenberg algebras:
\begin{equation}
\left[ a_{n},a_{n^{\prime }}\right] =n\delta _{n,n^{\prime }},\text{ \ \ \ }%
\left[ a_{n},\bar{a}_{n^{\prime }}\right] =0,\text{ \ \ }\left[ \bar{a}_{n},%
\bar{a}_{n^{\prime }}\right] =n\delta _{n,n^{\prime }}.
\end{equation}
The compactification condition implies that the spectrum of the momentum $p$%
\ is no-longer continuous but discrete with eigenvalues $n_{\phi }/R_{\phi }$
for $n_{\phi }\in \mathbb{Z}$. Finally, the Hamiltonian has the following
mode expansion:
\begin{equation}
E=\frac{2\pi v}{L}\left\{ \frac{1}{2}(a_{0}+\bar{a}_{0})+%
\sum_{n>0}a_{-n}a_{n}+\sum_{n>0}\bar{a}_{-n}\bar{a}_{n}\right\} .
\end{equation}

\section{$m$-reduction procedure on the plane}

The $m$-reduction technique \cite{VM} is at the basis of the derivation of
our Twisted Model (TM). Here, the main observation is that for any (\textit{%
mother}) CFT, defined by a given chiral algebra, there exists a class of
sub-theories (\textit{daughter theories}) parameterized by an integer $m$
with the same algebraic structure but different central charge $c_{m}=mc$.
The general characteristics of the daughter theory is the presence of
twisted boundary conditions (TBC) which are induced on the component fields
and are the signature of an interaction with a localized topological defect.

In this Section, contrary to our previous publications, we apply the $m$%
-reduction construction to scalar boson fields and not only to
their chiral components. That is done in order to consider carefully the action of the $m$%
-reduction on the zero modes which, as shown in the following
Section, will be crucial for the description of the
antiferromagnetic 2-leg\ spin-1/2 $XXZ$ ladder with Mobius
boundary conditions. In particular, we explicitly
describe the $m$-reduction only for the $m=2$ case when the \textit{mother%
} theory is the compactified boson theory ($c=1$), describing a single $%
XXZ$ spin-1/2 chain with general anisotropy parameter $\Delta $.

Let $\phi (z,\bar{z})$ be the compactified boson field of the
mother theory, we can define the following two scalar fields:
\begin{equation}
\widetilde{\mathcal{X}}(\text{w},\text{\={w}})=\frac{\phi (\text{w},\text{%
\={w}})+\phi (e^{i\pi }\text{w},e^{-i\pi }\text{\={w}})}{\sqrt{2}},\text{ \
\ \ }\widetilde{\Phi }(\text{w},\text{\={w}})=\frac{\phi (\text{w},\text{%
\={w}})-\phi (e^{i\pi }\text{w},e^{-i\pi }\text{\={w}})}{\sqrt{2}},
\label{Sym-AntiSym}
\end{equation}
which are respectively symmetric and antisymmetric under the action of the
generator $g_{2}$: $($w$,$\={w}$)\rightarrow (e^{i\pi }$w$,e^{-i\pi }$\={w}$%
) $ of the discrete group $Z_{2}$. The $m=2$-reduction is implemented by the
map $z=$w$^{2}$ which leads to the definition of the following \textit{%
daughter} scalar fields:
\begin{equation}
\mathcal{X}(z,\bar{z})=\widetilde{\mathcal{X}}(z^{1/2},\bar{z}^{1/2})\text{
\ \ and \ \ }\Phi (z,\bar{z})=\widetilde{\Phi }(z^{1/2},\bar{z}^{1/2})\text{.%
}
\end{equation}
The mode expansions of these last two fields are derived in terms of that of
$\phi $, Eq.(\ref{Boson-phi}), and read as:
\begin{equation}
\mathcal{X}(z,\bar{z})=q_{0}-i\alpha _{0}\ln z-i\bar{\alpha}_{0}\ln \bar{z}%
+\sum_{n\in \mathbb{Z}-\{0\}}\frac{i\alpha _{n}}{n}z^{-n}+\sum_{n\in \mathbb{%
Z}-\{0\}}\frac{i\bar{\alpha}_{n}}{n}\bar{z}^{-n},  \label{xx1}
\end{equation}
\begin{equation}
\Phi (z,\bar{z})=-\sqrt{2}\pi w+\sum_{n\in \mathbb{Z}}\frac{i\alpha _{n+1/2}%
}{n+1/2}z^{-(n+1/2)}+\sum_{n\in \mathbb{Z}}\frac{i\bar{\alpha}_{n+1/2}}{n+1/2%
}\bar{z}^{-(n+1/2)},  \label{xx2}
\end{equation}
where:
\begin{equation}
q_{0}=\sqrt{2}(q+\pi w),\text{ \ }\alpha _{n+l/2}=\frac{a_{2n+l}}{\sqrt{2}},%
\text{ \ }\bar{\alpha}_{n+l/2}=\frac{\bar{a}_{2n+l}}{\sqrt{2}}\text{ \ \
with }n\in N\text{\ and }l=0,1\text{,}
\end{equation}
and $w$ is the winding operator. The commutation relations of the \textit{%
daughter} modes $\left\{ q_{0},\{\alpha _{n+l/2}\},\{\bar{\alpha}%
_{n+l/2}\}\right\} $ follow from those of the \textit{mother} modes $\left\{
q,\{a_{n}\},\{\bar{a}_{n}\}\right\} $ and read as:
\begin{equation}
\left[ \alpha _{n+l/2},\alpha _{n^{\prime }+l^{\prime }/2}\right] =\left( n+%
\frac{l}{2}\right) \delta _{n,n^{\prime }}\delta _{l,l^{\prime }},\text{\ \ }%
\left[ \bar{\alpha}_{n+l/2},\bar{\alpha}_{_{n^{\prime }+l^{\prime }/2}}%
\right] =\left( n+\frac{l}{2}\right) \delta _{n,n^{\prime }}\delta
_{l,l^{\prime }},
\end{equation}
and
\begin{equation}
\left[ \alpha _{n+l/2},\bar{\alpha}_{_{n^{\prime }+l^{\prime }/2}}\right] =0,%
\text{\ \ }\left[ q_{0},\alpha _{n+l/2}\right] =i\delta _{0,n}\delta _{0,l},%
\text{ \ \ \ \ \ }\left[ q_{0},\bar{\alpha}_{n+l/2}\right] =i\delta
_{0,n}\delta _{0,l}.
\end{equation}
Let us point out that the modes $\{\alpha _{n}\}_{n\in \mathbb{Z}}$ and $\{%
\bar{\alpha}_{n}\}_{n\in \mathbb{Z}}$ of the field $\mathcal{X}(z,\bar{z})$
define two independent standard Heisenberg algebras while those of the field
$\Phi (z,\bar{z})$ ($\{\alpha _{n+1/2}\}_{n\in \mathbb{Z}}$ and $\{\bar{%
\alpha}_{n+1/2}\}_{n\in \mathbb{Z}}$) define two independent $Z_{2}$-\textit{%
twisted} Heisenberg algebras with \textit{half-integer} indices. The zero
mode analysis shows that the field $\mathcal{X}(z,\bar{z})$ is a
compactified boson with compactification radius $R_{\mathcal{X}}=R_{\phi }/%
\sqrt{2}$.

The description of the daughter theory proceeds now in a standard way. In
particular, for both the $Z_{2}$ \textit{untwisted} $\mathcal{X}$ and
\textit{twisted} $\Phi $ bosons we can define left and right chiral currents
and components of the stress-energy tensor. Let us consider explicitly only
the case of left chirality; the currents are:
\begin{equation}
J_{\mathcal{X}}(z)=i\partial _{z}\mathcal{X}(z,\bar{z})\left( =\sum_{n\in
\mathbb{Z}}\alpha _{n}z^{-n}\right) \text{, \ }J_{\Phi }(z)=i\partial
_{z}\Phi (z,\bar{z})\left( =\sum_{r\in \mathbb{Z}+1/2}\alpha
_{r}z^{-r}\right) ,
\end{equation}
and the corresponding components of the stress-energy tensor are:
\begin{equation}
T_{\mathcal{X}}(z)=-\frac{1}{2}\text{:}(\partial _{z}\mathcal{X})^{2}\text{:
, \ \ \ \ }T_{\Phi }(z)=-\frac{1}{2}\text{:}(\partial _{z}\Phi )^{2}\text{:}+%
\frac{1}{16z^{2}}.
\end{equation}
Notice that the second term in $T_{\Phi }(z)$ indicates that the $\Phi $%
-sector is built on the twisted vacuum generated by the left chiral twist
field $\sigma (z)$ with conformal dimension $1/16$. That is also manifest in
the mode expansion of the corresponding Virasoro generators:
\begin{equation}
L_{n}^{\mathcal{X}}=\frac{1}{2}\sum_{s\in \mathbb{Z}}\text{:}\alpha
_{n-s}\alpha _{s}\text{:},\text{ \ \ \ }L_{n}^{\Phi }=\frac{1}{2}\sum_{r\in
\mathbb{Z}+1/2}\text{:}\alpha _{n-r}\alpha _{r}\text{:}+\frac{\delta _{0,n}}{%
16}\text{ \ with }n\in \mathbb{Z.}
\end{equation}
Let us stress that $\{L_{n}^{\Phi }\}$ and $\{L_{n}^{\mathcal{X}}\}$ define
two independent Virasoro algebras both with central charge $c=1$. Thus the
daughter theory has the left component of the stress-energy tensor equal to $%
T(z)=T_{\mathcal{X}}(z)+T_{\Phi }(z)$, which defines a Virasoro algebra with
central charge $c=2$ ($m=2$-reduction).

\section{2-leg $XXZ$ spin-1/2 ladders with Mobius boundary conditions: the
continuum limit}

In this Section we show how the TM, generated by $m=2$-reduction,
describes well the 2-leg $XXZ$ spin-1/2 ladder with general
anisotropy parameter $\Delta $ arranged in a closed geometry for
PBC and MBC boundary conditions. The TM is characterized by two
topological sectors, the untwisted and the twisted ones, which
describe the continuum limit of the 2-leg $XXZ$ spin-1/2 ladder
with PBC and MBC respectively. In order to establish that it is
enough to show that the spin-1/2 $XXZ$ ladder with MBC in the
continuum is naturally mapped in the twisted sector of our TM,
i.e. the sector of TM generated by the $m=2$-reduction. Indeed the
untwisted sector of TM can be described in terms of two untwisted
boson fields, with the appropriate compactification radius, which
define the continuum limit of the 2-leg $XXZ$ spin-1/2 ladder with
PBC. Finally, the special value of the anisotropy parameter
$\Delta =1$ is considered and the symmetry properties of the
corresponding physical system briefly underlined.

Let us recall that the 2-leg $XXZ$ spin-1/2 ladder with $M$ odd sites and
MBC coincides with a system of two $XXZ$ spin-1/2 chains, each one with $%
(M+1)/2$ sites, which are closed with periodicity condition in one \textit{%
gluing} site which is common to the two chains. This induces a local
deformation of the interaction which has a purely topological nature
(topological defect) due to the fact that the gluing site has four nearest
neighboring sites (two for each chain) with which it interacts. The presence
of the topological defect implies that the ground state of the system in the
thermodynamical limit is not in the sector $S_{tot}^{z}=0$. Indeed, while
the full system has \textit{even} size $L_{2}=2L$ ($L=a(M+1)/2$ being the
size of the single chain and $a$ the lattice spacing) the number of quantum
sites is the \textit{odd} integer $M$. We can perform now a bosonization
analysis for each chain of the system, as shown in Section 3. In this way we
obtain, for the up and down chain respectively, two boson fields $\phi _{Up}$
and $\phi _{Dw}$ compactified on the two circles (up and down) of the same
length $L$ and with the same compactification radius $R_{\phi }=\sqrt{%
2\left( \pi -\arccos \Delta \right) /\pi }$. Here, the presence of the
topological defect implies the following boundary condition for the fields
in the gluing point ($x_{Up}=x_{Dw}=0$):
\begin{equation}
\phi _{Up}(0,t)=\phi _{Dw}(0,t),  \label{Boundary-Up/Dw}
\end{equation}
which in turn allows us to define the following folding field $\phi $ on the
full system:
\begin{equation}
\phi (x,t)=\left\{
\begin{array}{l}
\phi _{Dw}(x,t),\text{ \ \ for }0\leq x\leq L, \\
\\
\phi _{Up}(x-L,t),\text{ \ \ for }L\leq x\leq 2L.
\end{array}
\right.
\end{equation}
The non-trivial topology of the whole system is reflected into the fact that
the points $x=0$ and $x=L$ coincide; that is, the compactification space of
the field $\phi $ is not a circle of length $L_{2}$ but an \textit{eight} of
same length. It is now simple to show that on this \textit{eight}-shaped
space the field $\phi $ has compactification radius $R_{\phi }$. To this aim
it is enough to observe that, on the up/down circle, starting from a given
point $x_{Up/Dw}$ we come back to it under a shift of $L$ while, on the
\textit{eight,} we need a shift of $L_{2}=2L$, so that, it results\footnote{%
Let us notice that the boundary condition (\ref{Boundary-Up/Dw}) naturally
leads to assume that the fields $\phi _{Dw}$ and $\phi _{Up}$ are taken in
the same \textit{winding sector}, that is $w_{\phi _{Dw}}=w_{\phi _{Up}}$,
so that:
\begin{equation*}
m_{\phi _{Dw}}=m_{\phi _{Up}}=m_{\phi }
\end{equation*}
}:
\begin{equation*}
\phi (x+2L,t)=\left\{
\begin{array}{l}
\phi _{Dw}(x+L,t),\text{ \ \ for }0\leq x\leq L, \\
\\
\phi _{Up}(x,t),\text{ \ \ for }L\leq x\leq 2L,
\end{array}
\right.
\end{equation*}
\begin{equation*}
\text{\ \ \ \ \ \ \ \ \ \ \ \ \ \ \ \ \ \ \ \ \ \ \ \ \ \ \ \ \ \ \ \ \ }%
=\left\{
\begin{array}{l}
\phi _{Dw}(x,t)+2\pi m_{\phi }R_{\phi },\text{ \ \ for }0\leq x\leq L, \\
\\
\phi _{Up}(x-L,t)+2\pi m_{\phi }R_{\phi },\text{ \ \ for }L\leq x\leq 2L,
\end{array}
\right.
\end{equation*}
that is:
\begin{equation}
\phi (x+2L,t)=\phi (x,t)+2\pi m_{\phi }R_{\phi }\text{.}
\end{equation}
Now we can introduce the complex coordinates:
\begin{equation}
\text{w}=e^{2\pi (\tau -ix)/L_{2}},\text{ \ \ \ \ \ \={w}}=e^{2\pi (\tau
+ix)/L_{2}}
\end{equation}
where $\tau =ivt$ (Euclidean space-time) and define the fields $\widetilde{%
\mathcal{X}}($w$,$\={w}$)$ and $\widetilde{\Phi }($w$,$\={w}$)$ according to
the formula (\ref{Sym-AntiSym}). Let us observe that on the square covering
plane:
\begin{equation}
z=\text{w}^{2}=e^{2\pi (\tau -ix)/L},\text{ \ \ \ \ \ }\bar{z}=\text{\={w}}%
^{2}=e^{2\pi (\tau +ix)/L}
\end{equation}
the coincidence of the points $x=0$ and $x=L$, required by the \textit{eight
shape }of the space in the $x$-coordinate, now holds automatically. In fact,
the space in the $x$-coordinate is now a circle of length $L$. The above
covering is nothing else but the map defined in the previous Section, which
implements the $m=2$-reduction on the field $\phi ($w$,$\={w}$)$ producing
the two independent boson fields $\mathcal{X}(z,\bar{z})$ and $\Phi (z,\bar{z%
})$ of the daughter theory.

Summarizing, by using the $m=2$-reduction technique, we have transformed the
boson field $\phi $ compactified on the \textit{eight}-shape space of length
$2L $ with compactification radius $R_{\phi }$ into the two independent
boson fields $\mathcal{X}$ and $\Phi $ compactified on a circle of length $L$%
, where $R_{\mathcal{X}}=R_{\phi }/\sqrt{2}$ is the compactification radius
of $\mathcal{X}$. Thus the 2-leg $XXZ$ spin-1/2 ladder with general
anisotropy parameter $\Delta $ and Mobius boundary conditions results
described by the twisted sector of our TM.

Let us focus on the isotropic case, that is a $XXX$ ladder, corresponding to
the value $\Delta =1$ of the anisotropy parameter. The mother theory now is
a theory of a free massless boson field $\phi $ with central charge $c=1$,
compactified on a circle with compactification radius $R_{\phi }^{2}=\frac{1%
}{K}=2$. As a result of the $2$-reduction\ procedure we obtain a $c=2$
theory of two boson fields, Eqs. (\ref{xx1}) and (\ref{xx2}), which describe
the spin chains of the two legs. Such a daughter theory is an orbifold one,
which decomposes into a tensor product of two CFTs, a twisted invariant one
with $c=3/2$, realized by the boson $\mathcal{X}(z,\bar{z})$ and a Ramond
Majorana fermion,\ while the second one has $c=1/2$ and is realized in terms
of a Neveu-Schwarz Majorana fermion. Such a factorization, $\widehat{su(2)}%
_{2}\otimes ($Ising$)$, is evident in the modular invariant partition
function within the torus topology (\ref{partition1}) (see Appendix A).

Let us notice that the non abelian bosonization model for the weakly coupled
frustrated $XXX$ ladder, introduced in Ref. \cite{allen}, with symmetry $%
\left( \text{Ising}\right) ^{4}$ corresponds to only one topological sector
of our orbifold theory, the untwisted one, which describes PBC imposed at
the ends of the two spin-1/2 chains. In closed geometries, nevertheless, the
appropriate theory has to include all the relevant boundary conditions, as
MBC ones, and this is done by our TM due to the presence of the extra
twisted sector. The different boundary conditions correspond to boundary
states at the end of the finite ladder, which are codified in terms of the
different sectors in the modular invariant partition function. The
decomposition into two topological sectors just outlined gives rise to an
essential difference in the low energy spectrum for the system under study.
Indeed, in the untwisted sector (PBC) only two-spinon excitations are
possible (i.e. $SU(2)$ integer spin representations) while, in the twisted
sector (MBC) the presence of a topological defect provides a clear evidence
of single-spinon excitations (i.e. $SU(2)$ half-integer spin
representations). All that takes place in close analogy with fully
frustrated Josephson ladders in the extremely quantum limit which are
expected to map to spin-1/2 zigzag ladders \cite{nishi}.

\section{Renormalization group analysis}

In this Section we deal with the weakly interacting 2 leg\ spin-1/2 ladder
system in the continuum by turning on the four different perturbations (\ref
{p1})-(\ref{p3}) and (\ref{p4}). We restrict our analysis to the isotropic
case $\Delta =1$, and study the different renormalization group (RG)
trajectories flowing from the UV fixed point, described by our TM.

It is worth recalling that our TM is a $Z_{2}$-orbifold with central charge $%
c=2$ and that the model described by standard bosonization
technique coincides just with the untwisted sector of TM. In the
next two subsections, we extend to the twisted sector of TM the RG
analysis done previously in
\cite{shelton1},\cite{wang1} and \cite{allen} for the perturbations (\ref{p1}%
), (\ref{p2}) and (\ref{p3}). Moreover, in subsection 6.2, we
introduce the new interaction (\ref{p4}) within its fermionic
representation and study the RG flow generated by the combined
action of the perturbations (\ref{p2}) and (\ref{p4}). Depending
on the perturbing term, we obtain different infrared (IR) fixed
points corresponding to different physical behaviors.

The perturbing terms are scalar ones and get expressed in terms of our
bosonic daughter fields $\mathcal{X}(z,\bar{z})$ and $\Phi (z,\bar{z})$ and
the corresponding dual fields $\mathcal{Y}(z,\bar{z})$ and $\Theta (z,\bar{z%
})$, which admit the following representation by left and right chiral
components:
\begin{equation}
\mathcal{X}(z,\bar{z})=-iw_{\mathcal{X}}\ln \frac{z}{\bar{z}}+X(z)+\bar{X}(%
\bar{z}),\text{ \ \ }\mathcal{Y}(z,\bar{z})=X(z)-\bar{X}(\bar{z}),
\label{scalar1}
\end{equation}
and
\begin{equation}
\Phi (z,\bar{z})=-2\pi w_{\mathcal{X}}+\varphi (z)+\bar{\varphi}(\bar{z}),%
\text{ \ \ }\Theta (z,\bar{z})=\varphi (z)-\bar{\varphi}(\bar{z}),
\label{scalar2}
\end{equation}
where $w_{\mathcal{X}}$ is the winding operator with eigenvalues $m_{%
\mathcal{X}}R_{\mathcal{X}}/2$ and $m_{\mathcal{X}}\in \mathbb{Z}$. In the
following a representation in terms of four Majorana fermion fields turns
out useful, whose holomorphic components ${(1/2,0)}$ are defined as:{\
\begin{equation}
\psi _{1}(z)=\text{sin}X\left( z\right) ,\text{ }\psi _{2}(z)=\text{cos}%
X\left( z\right) ,\text{ }\psi _{3}(z)=\frac{\text{sin}\varphi \left(
z\right) }{\sqrt{z}},\text{ }\psi _{0}(z)=\frac{\text{cos}\varphi \left(
z\right) }{\sqrt{z}},  \label{fermion1}
\end{equation}
while the anti-holomorphic ones }${(0,1/2)}${\ are:
\begin{equation}
\bar{\psi}_{1}(\bar{z})=-\text{sin}\bar{X}\left( \bar{z}\right) ,\text{ }%
\bar{\psi}_{2}(\bar{z})=\text{cos}\bar{X}\left( \bar{z}\right) ,\text{ }\bar{%
\psi}_{3}(\bar{z})=-\frac{\text{sin}\bar{\varphi}\left( \bar{z}\right) }{%
\sqrt{\bar{z}}},\text{ }\bar{\psi}_{0}(\bar{z})=\frac{\text{cos}\bar{\varphi}%
\left( \bar{z}\right) }{\sqrt{\bar{z}}}.  \label{fermion4}
\end{equation}
}In the present case ($\Delta =1$), the TM is a $Z_{2}$-{orbifold with }$%
\widehat{su(2)}_{2}\otimes \bar{I}$ symmetry, where the Ising factor $\bar{%
I}$ (which is $Z_{2}$-antisymmetric) is generated\footnote{%
The field $\psi _{0}(z)$ has antiperiodic boundary conditions on the plane:
\begin{equation*}
\psi _{0}(e^{2i\pi }z)=-\psi _{0}(z).
\end{equation*}
The fields $\psi _{i}(z)$ ($i=1,2,3$) have periodic boundary conditions on
the plane:
\begin{equation*}
\psi _{i}(e^{2i\pi }z)=\psi _{i}(z).
\end{equation*}
} by $\psi _{0}(z)$ (and the corresponding anti-holomorphic component $\bar{%
\psi}_{0}(\bar{z})$), and the $\widehat{su(2)}_{2}$ factor is generated by
the currents:
\begin{equation}
J^{\pm }(z)=\psi _{3}(z)e^{\pm iX(z)},\text{ }J^{3}(z)=i\partial X(z),
\label{currents1}
\end{equation}
and analogous expressions hold for the anti-holomorphic components. The
Lagrangian describing the UV fixed point for the $XXX$ ladder is:
\begin{equation}
\mathcal{L}_{0}=\frac{1}{8\pi }(\partial _{\mu }\mathcal{X}\partial ^{\mu }%
\mathcal{X}+\partial _{\mu }\Phi \partial ^{\mu }\Phi )  \label{lagrangian11}
\end{equation}
while the perturbing terms $V$ depend on the particular system under study:
\begin{equation}
\mathcal{L}=\mathcal{L}_{0}-V.  \label{lagrangian111}
\end{equation}

\subsection{Ladder and 4-Spin perturbations: massive flow}

In the following we deal with the interacting terms $H_{\text{Railroad}}$, $%
H_{\text{4-Spin}}$, Eqs. (\ref{p1}) and (\ref{p3}), which we write
in the continuum limit for the twisted sector of TM, retaining
only the relevant terms.

In the continuum the railroad interaction\footnote{%
From here on we denote with $V_{\text{x}}$ the Lagrangian density which
defines the continuum limit of the lattice interaction $H_{\text{x}}$, with
x=Railroad, Zigzag, 4-Spin, 4-Dimer.} gives rise to relevant perturbations
(with conformal dimensions $(1/2,1/2)$) to the UV fixed point, described by
our TM theory with central charge $c=2$. Such massive terms are given in
terms of the boson fields (\ref{scalar1})-(\ref{scalar2}) by:
\begin{equation}
V_{\text{Railroad}}=-m_{\text{R}}(\text{cos}\mathcal{X}(z,\bar{z})-\text{cos}%
\Phi (z,\bar{z})+2\text{cos}\Theta (z,\bar{z})),  \label{per1}
\end{equation}
where $m_{\text{R}}\varpropto J_{\perp }^{R}$, \ $J_{\perp }^{R}$ being the
coupling constant of the railroad interaction (\ref{p1}). In the Majorana
fermion representation we get:
\begin{equation}
V_{\text{Railroad}}=-im_{\text{R}}\sum\limits_{i=1}^{3}\psi _{i}(z)\bar{\psi}%
_{i}(\bar{z})+3im_{\text{R}}\psi _{0}(z)\bar{\psi}_{0}(\bar{z}),
\label{per2}
\end{equation}
where it appears clearly that the $\psi _{i}$ ($i=1,2,3$) fields form an
Ising triplet with the same mass $m_{\text{R}}$ (the $\widehat{su(2)}_{2}$
sector) and the remaining field is an Ising singlet (the $\bar{I}$ sector)
with a larger mass $-3m_{\text{R}}$ \cite{shelton1}.

Let us now switch on the 4-spin interaction $H_{\text{4-Spin}}$, Eq. (\ref
{p3}), first introduced in Ref. \cite{wang1}, whose expression in the
continuum limit can be found starting from the continuum limit for the
dimerization operators:
\begin{equation}
\epsilon _{+}\sim \mu _{1}\mu _{2}\mu _{3}\sigma _{0},\ \ \ \ \epsilon
_{-}\sim \sigma _{1}\sigma _{2}\sigma _{3}\mu _{0},  \label{per3}
\end{equation}
where $\epsilon _{\pm }=\epsilon _{Up}\pm \epsilon _{Dw}$ and
$\epsilon _{Up/Dw}$ are the continuum analogues in the fermionic
representation of the spin-field dimerizations corresponding to
the up and down leg respectively, quoted in Eq. (\ref{p3a}). Now,
by using the fusion rules of the Ising model
and dropping the most singular term, we can regularize the OPE lim$%
_{z\rightarrow w}:\left( \epsilon _{+}(w,\bar{w})\epsilon _{+}(z,\bar{z}%
)-\epsilon _{-}(w,\bar{w})\epsilon _{-}(z,\bar{z})\right) :$ and write the
continuum limit of the interaction $H_{\text{4-Spin}}$ as:
\begin{equation}
V_{\text{4-Spin}}(z,\bar{z})=im_{\epsilon }\sum\limits_{i=0}^{3}\psi _{i}(z)%
\bar{\psi}_{i}(\bar{z}).  \label{per4}
\end{equation}
Such a relevant mass term (with conformal dimensions $(1/2,1/2)$) can be
rephrased in boson language as:
\begin{equation}
V_{\text{4-Spin}}=m_{\epsilon }(\text{cos}\mathcal{X}(z,\bar{z})+\text{cos}%
\Phi (z,\bar{z})),  \label{per5}
\end{equation}
where $m_{\epsilon }\varpropto J_{\epsilon }$, $J_{\epsilon }$ being the
coupling constant of the 4-spin interaction $H_{\text{4-Spin}}$. As we see
from Eq. (\ref{per4}), such a case is different from the railroad
interaction (\ref{per2}) in that it gives rise to the same mass contribution
$m_{\epsilon }$ for all the Ising fields $\psi _{i}$ ($i=0,1,2,3$). That
allows the triplet or singlet mass to vanish also for finite (different from
zero) values of the coupling constants $J_{\bot }^{R},J_{\epsilon }$, i.e.
far from the UV conformal fixed point\footnote{%
The whole theory $\widehat{su(2)}_{2}\bigotimes \bar{I}$ is not a conformal
one because one of its two factors is always massive in each of the two
possible RG flows: ($m_{t}=0,m_{s}\neq 0$) and ($m_{t}\neq 0,m_{s}=0$).} $%
c=2 $. \ Then two possible trajectories in the RG flow arise, the first
characterized by ($m_{t}=0,m_{s}\neq 0$) and the second by ($m_{t}\neq
0,m_{s}=0$), where the mass different form zero increases more and more
along the flow, while the vanishing one remains unchanged. In such a case
the low energy spectrum is well described by an effective conformal field
theory which is obtained after the decoupling of the massive component. We
get two possible RG flows: a flow to an IR fixed point with central charge $%
c=3/2$ as a result of the Ising $\bar{I}$ decoupling and a flow to a
different IR fixed point with central charge $c=1/2$ as a result of the $%
\widehat{su(2)}_{2}$ decoupling.

Let us now analyze the behavior of the staggered magnetization operator $%
n^{\pm }$ as well as the dimerization one $\epsilon _{\pm }$ along the
massive RG flow just discussed. Within the $c=2$ theory they can be
expressed in terms of the boson fields (\ref{scalar1})-(\ref{scalar2}) as:
\begin{eqnarray}
n^{+} &\sim &(\text{cos}\frac{\mathcal{Y}(z,\bar{z})}{2}\text{cos}\frac{%
\Theta (z,\bar{z})}{2},\text{ sin}\frac{\mathcal{Y}(z,\bar{z})}{2}\text{cos}%
\frac{\Theta (z,\bar{z})}{2},\text{ sin}\frac{\mathcal{X}(z,\bar{z})}{2}%
\text{cos}\frac{\Phi (z,\bar{z})}{2}),  \label{per6} \\
&&  \notag \\
n^{-} &\sim &(\text{sin}\frac{\mathcal{Y}(z,\bar{z})}{2}\text{sin}\frac{%
\Theta (z,\bar{z})}{2},\text{ cos}\frac{\mathcal{Y}(z,\bar{z})}{2}\text{sin}%
\frac{\Theta (z,\bar{z})}{2},\text{ cos}\frac{\mathcal{X}(z,\bar{z})}{2}%
\text{sin}\frac{\Phi (z,\bar{z})}{2}),  \label{per7} \\
&&  \notag \\
\epsilon _{+} &\sim &\text{cos}\frac{\mathcal{X}(z,\bar{z})}{2}\text{cos}%
\frac{\Theta (z,\bar{z})}{2},\ \ \ \ \ \ \epsilon _{-}\sim \text{sin}\frac{%
\mathcal{X}(z,\bar{z})}{2}\text{sin}\frac{\Theta (z,\bar{z})}{2}.
\label{per8}
\end{eqnarray}
Indeed in the fermionic representation they are written as:
\begin{eqnarray}
n^{+} &\sim &(\sigma _{1}\mu _{2}\sigma _{3}\sigma _{0},\mu _{1}\sigma
_{2}\sigma _{3}\sigma _{0},\sigma _{1}\sigma _{2}\mu _{3}\sigma _{0}),\text{
\ \ }\epsilon _{+}\sim \mu _{1}\mu _{2}\mu _{3}\sigma _{0},  \label{per9} \\
&&  \notag \\
n^{-} &\sim &(\mu _{1}\sigma _{2}\mu _{3}\mu _{0},\sigma _{1}\mu _{2}\mu
_{3}\mu _{0},\mu _{1}\mu _{2}\sigma _{3}\mu _{0}),\text{ \ \ }\epsilon
_{-}\sim \sigma _{1}\sigma _{2}\sigma _{3}\mu _{0}.  \label{per11}
\end{eqnarray}
In the disordered $\bar{I}$ Ising singlet phase characterized by $m_{s}>0$ ($%
m_{t}=0$) we get:
\begin{equation}
\langle \sigma _{0}\rangle =0,\text{ \ \ \ }\langle \mu _{0}\rangle \neq 0.
\label{per12}
\end{equation}
As a consequence the fields $\epsilon _{-}$ and $n^{-}$ can be expressed in
terms of the primary fields of the CFT with $\widehat{su(2)}_{2}$ symmetry
as:
\begin{equation}
n^{-}\sim (\mu _{1}\sigma _{2}\mu _{3},\sigma _{1}\mu _{2}\mu _{3},\mu
_{1}\mu _{2}\sigma _{3}),\ \ \ \ \epsilon _{-}\sim \sigma _{1}\sigma
_{2}\sigma _{3},  \label{per13}
\end{equation}
and the corresponding correlation functions of the relative staggered
magnetization and dimerization field behave as:
\begin{equation}
\langle n^{-}(z)n^{-}(0)\rangle \sim \langle \epsilon _{-}(z)\epsilon
_{-}(0)\rangle \sim \left| z\right| ^{-3/4}.  \label{per14}
\end{equation}
Thus, the critical point $m_{t}=0$ is an IR fixed point with central charge $%
c=\frac{3}{2}$.

Likewise in the disordered $\widehat{su(2)}_{2}$ Ising triplet phase
characterized by $m_{t}>0$ ($m_{s}=0$) we get:
\begin{equation}
\langle \sigma _{i}\rangle =0,\text{ \ \ \ }\langle \mu _{i}\rangle \neq 0%
\text{ \ \ }\forall i\in \{1,2,3\}.  \label{per15}
\end{equation}
As a consequence the field $\epsilon _{+}$ can be expressed in terms of the
following primary field within the $\bar{I}$ CFT:
\begin{equation}
\epsilon _{+}\sim \sigma _{0},  \label{per16}
\end{equation}
and the corresponding correlation function follows a power law:
\begin{equation}
\langle \epsilon _{+}(z)\epsilon _{+}(0)\rangle \sim \left|z\right|^{-1/4}.
\label{per17}
\end{equation}
Thus the critical point $m_{s}=0$ belongs to the Ising universality class
with central charge $c=\frac{1}{2}$ and signals a transition to a
spontaneously dimerized phase with $\langle \epsilon _{+}\rangle \neq 0$.

\subsection{Zigzag and 4-Dimer perturbations: massive flow}

In the following we switch off the perturbing terms $H_{\text{Ladder}}$, $H_{%
\text{4-Spin}}$ and deal with the interacting terms $H_{\text{ZigZag}}$, $H_{%
\text{4-Dimer}}$, Eqs. (\ref{p2}) and (\ref{p4}), which we write
in the continuum limit for the twisted sector of TM. Let us recall
that the system of two spin-1/2 chains coupled via a zigzag
interaction, with Hamiltonian $H_{0}+$ $H_{\text{ZigZag}}$, has
been widely investigated in the literature
\cite{white1}\cite{allen}\cite{chiral}\cite {itoi} in the
weak-coupling regime $J_{0}\gg J_{\perp }^{Z}$ via a non abelian
bosonization approach\cite{nbos}. In this limit of two
quasi-decoupled chains with periodic boundary conditions, the
system in the continuum has been described in terms of two level-1
Wess-Zumino-Witten (WZW) field theories or equivalently in terms
of four Majorana fermions coupled by some perturbations. As a
first step toward the RG analysis of the system in the presence of
both the interactions (\ref{p2}) and (\ref{p4}), let us start
extending the RG analysis of the system with only the zigzag
interaction in the twisted sector of our TM. In the present case,
the implementation of the continuum limit has to be brought out
carefully, being eventually the zigzag interaction described by a
marginal operator. The continuum limit of the Hamiltonian $H_{0}$
of two non-interacting chains leads to the free-fermion
Lagrangian:
\begin{equation}
\mathcal{L}_{0}=\frac{1}{2\pi }\sum_{i=0}^{3}v_{i}\left( \psi _{i}\overline{%
\partial }\psi _{i}+\overline{\psi }_{i}\partial \overline{\psi }_{i}\right)
\label{alsen2}
\end{equation}
where $v_{0}=...=v_{3}=v\sim J_{0}a$ is the velocity of spin excitation in
isolated chains, $a$ being the lattice spacing, plus a marginal interaction
\cite{allen}:
\begin{equation}
V_{U}=-\lambda _{U}(O_{1}+O_{2}),  \label{per18}
\end{equation}
where $\lambda _{U}\sim U/|t|\geq 0$ and:
\begin{eqnarray}
O_{1} &=&\psi _{1}(z)\bar{\psi}_{1}(\bar{z})\psi _{2}(z)\bar{\psi}_{2}(\bar{z%
})+\psi _{1}(z)\bar{\psi}_{1}(\bar{z})\psi _{3}(z)\bar{\psi}_{3}(\bar{z}%
)+\psi _{2}(z)\bar{\psi}_{2}(\bar{z})\psi _{3}(z)\bar{\psi}_{3}(\bar{z}),
\label{per19} \\
&&  \notag \\
O_{2} &=&\psi _{0}(z)\bar{\psi}_{0}(\bar{z})(\psi _{1}(z)\bar{\psi}_{1}(\bar{%
z})+\psi _{2}(z)\bar{\psi}_{2}(\bar{z})+\psi _{3}(z)\bar{\psi}_{3}(\bar{z})),
\label{per20}
\end{eqnarray}
within the Majorana fermion representation. The zigzag interaction $H_{\text{%
ZigZag}}$ has instead the following form:
\begin{equation}
V_{\text{ZigZag}}=\lambda _{\text{Z}}[(O_{1}-O_{2})+%
\sum_{i=0}^{3}(T^{(i)}(z)+\bar{T}^{(i)}(\bar{z}))],  \label{per21}
\end{equation}
where $\lambda _{\text{Z}}\sim J_{\perp }^{Z}/|t|\geq 0$ and $J_{\perp
}^{Z}\geq 0$. Such an marginal interaction contains a non scalar term given
by the sum over the energy-momentum tensor of the fermionic fields, whose
effect is simply that of renormalizing the fields and the velocities:
\begin{eqnarray}
&&
\begin{array}{cc}
\left( \psi _{i},\overline{\psi }_{i}\right) \rightarrow \frac{1}{\sqrt{%
1+\pi \lambda _{\text{Z}}}}\left( \psi _{i},\overline{\psi }_{i}\right)
\mathcal{,} & i=0,1,2,3
\end{array}
\\
&&
\begin{array}{cc}
v_{0}\rightarrow v_{0}\frac{1+3\pi \lambda _{\text{Z}}}{1-3\pi \lambda _{%
\text{Z}}}, & v_{i}\rightarrow v_{i}\frac{1-\pi \lambda _{\text{Z}}}{1+\pi
\lambda _{\text{Z}}},i=1,2,3
\end{array}
.
\end{eqnarray}
After renormalization the whole effect of the perturbing terms $V_{\text{%
ZigZag}}$ and $V_{U}$ can be expressed as a marginal interaction:
\begin{equation}
\mathcal{V}_{\text{ZigZag}}=\lambda _{+}^{0}(O_{1}+O_{2})+\lambda
_{-}^{0}(O_{1}-O_{2}),  \label{per22}
\end{equation}
where $\lambda _{\pm }^{0}=\frac{1}{2\left( 1+\pi \lambda _{\text{Z}}\right)
}\left\{ \mp \frac{\lambda _{\text{Z}}+\lambda _{U}}{1-3\pi \lambda _{\text{Z%
}}}+\frac{\lambda _{\text{Z}}-\lambda _{U}}{1+\pi \lambda _{\text{Z}}}%
\right\} $ and satisfy the following RG equations \cite{allen}:
\begin{equation}
\frac{d\lambda _{\pm }^{0}}{d\ln L}=8\pi \left( \lambda _{\pm }^{0}\right)
^{2}.  \label{per23}
\end{equation}
The value of the parameters $\lambda _{U}\geq 0$ and $\lambda _{\text{Z}%
}\geq 0$ gives rise to $\lambda _{+}^{0}<0$ and $\lambda _{-}^{0}$\ small
and positive. Thus, under the flow (\ref{per23}) $\lambda _{+}^{0}$
renormalizes to zero ($\lambda _{+}^{0}\rightarrow 0^{-}$) while $\lambda
_{-}^{0}$\ increases being marginally relevant and that results in a
dynamical length scale $\xi \sim e^{1/\lambda _{-}^{0}}$. A mass scale
appears dynamically and provides a non vanishing mass for the four fermions:
$m_{i}\sim v_{i}\xi ^{-1}$, $i=1,2,3$, $m_{0}\sim v_{0}\xi ^{-1}$, with $%
m_{1}=m_{2}=m_{3}=m>0$, $m_{0}<-m$.

In such a picture it is not possible to extract trajectories in
the RG flow characterized by a vanishing mass in the $c=1/2$ or
$c=3/2$ sector respectively. In order to obtain such trajectories
we need to introduce a new perturbing term, the 4-dimer
$H_{\text{4-Dimer}}$, defined in Eq. (\ref {p4}). In the continuum
the double dimerization operator $\varkappa _{i}=\epsilon
_{i}^{Dw}\epsilon _{i}^{Up}$ can be defined by regularizing the
following expression:
\begin{equation}
\varkappa (z,\bar{z})=\frac{C_{\varkappa }}{4}\text{ lim}_{z\rightarrow
w}:\left( \epsilon _{+}(w,\bar{w})\epsilon _{+}(z,\bar{z})-\epsilon _{-}(w,%
\bar{w})\epsilon _{-}(z,\bar{z})\right) :,  \label{per24}
\end{equation}
as done in the previous Subsection, Eq. (\ref{per4}); the net result is:
\begin{equation}
\varkappa (z,\bar{z})=\sum\limits_{i=0}^{3}\psi _{i}(z)\bar{\psi}_{i}(\bar{z}%
)  \label{per25}
\end{equation}
The continuum limit of the interaction $H_{\text{4-Dimer}}$:
\begin{equation}
V_{\text{4-Dimer}}=\lambda _{\varkappa }(O_{1}+O_{2})=\lambda _{\varkappa
}\sum\limits_{i\neq j=0}^{3}\psi _{j}(z)\bar{\psi}_{j}(\bar{z})\psi _{i}(z)%
\bar{\psi}_{i}(\bar{z}),  \label{per26}
\end{equation}
where $\lambda _{\varkappa }\sim J_{\varkappa }$, is now obtained applying
once again the regularization procedure to the OPE lim$_{z\rightarrow
w}:\left( \varkappa (w,\bar{w})\varkappa (z,\bar{z})\right) :$; the whole
perturbation turns out to be:
\begin{equation}
\mathcal{V}_{\text{Tot}}\equiv \mathcal{V}_{\text{ZigZag}}+V_{\text{4-Dimer}%
}=\lambda _{+}(O_{1}+O_{2})+\lambda _{-}(O_{1}-O_{2})  \label{per27}
\end{equation}
where $\lambda _{+}=\lambda _{+}^{0}+\lambda _{\varkappa }$ and $\lambda
_{-}=\lambda _{-}^{0}$. Such a general interaction contains two parameters, $%
\lambda _{+}$ and $\lambda _{-}$, which can be changed independently by
changing the coupling constants ($J_{\perp }^{Z}$,$J_{\varkappa }$). In
particular, that allows us to define a path in the RG flow characterized by
a vanishing singlet mass, i.e. ($m_{t}\neq 0,m_{s}=0$). In order to obtain
such a path let us rewrite the general perturbation (\ref{per27}) as:
\begin{equation}
\mathcal{V}_{\text{Tot}}=\Lambda _{1}O_{1}+\Lambda _{2}O_{2}  \label{per28}
\end{equation}
where $\Lambda _{1}=\lambda _{+}+\lambda _{-}$ and $\Lambda
_{2}=\lambda _{+}-\lambda _{-}$. Notice that a mass $m_{s}$ for
the singlet could be provided dynamically only by means of the
interaction $O_{2}$, because of the presence of the term $\psi
_{0}(z)\bar{\psi}_{0}(\bar{z})$. The vanishing of the singlet mass
$m_{s}$ can then be obtained only by requiring the marginality of
the operator $O_{2}$ along the RG flow. This selects out the
condition $\lambda _{+}=\lambda _{-}$ which makes $\Lambda _{2}$
to vanish and which is left invariant under the RG equations.
Within a self-consistent mean field approximation (see Appendix
B), it is possible to show that along this RG flow trajectory the
singlet mass $m_{s}$ indeed vanishes while the triplet mass
$m_{t}$ is dynamically generated and reads:
\begin{equation}
m_{t}\sim \pm v_{t}\digamma \exp (-1/8\pi \lambda _{+}),  \label{per29}
\end{equation}
where $\digamma $ is a momentum cutoff and $v_{t}$ is the spin triplet
velocity. The above analysis shows that, under the introduction of the
4-Dimer interaction, we are able to describe an RG trajectory flowing from
our TM, the $c=2$ UV$\ $fixed point, toward the Ising $\bar{I}$, the $c=1/2$
IR fixed point, as a result of the dynamical generation only of the triplet
mass $m_{t}$ and the consequent decoupling of $\widehat{su(2)}_{2}$.

It is worth noting that the four interactions above introduced could not
produce a massless flow to a conformal IR fixed point starting from the UV
one with $c=2$. To this aim further interactions which could compete with
the mass generation are needed. Let us mention here that following the same
analysis developed in \cite{noi}\ we have that in the continuum limit the
perturbation:
\begin{equation}
V_{3/2}=\sigma _{0}\sigma _{3},  \label{per34}
\end{equation}
gives rise to a massless RG flow to the $c=3/2$ IR fixed point, $\widehat{%
su(2)}_{2}$. The perturbing terms which induce in the continuum limit
massless flows to the $c=1$ and $c=1/2$ fixed points can be similarly found
and their lattice realization is under analysis.

\section{Conclusions and outlooks}

In this paper a complete CFT for antiferromagnetic spin-1/2 2-leg $XXZ$
ladders with general anisotropy parameter $\Delta $ and arranged in a closed
geometry has been developed. Two kinds of boundary conditions on gluing the
ends of the two legs have been considered, that is PBC and MBC. The two
topologically inequivalent configurations which arise for an even and an odd
number of sites in the ladder system are deeply investigated and the
implications of the presence of a topological defect on the spectrum and the
low-energy excitations are exploited. The net result in the continuum limit
is a $Z_{2}$-orbifold CFT with central charge $c=2$, obtained through $2$%
-reduction on the single $XXZ$ antiferromagnetic spin-1/2 chain,
whose topological sectors account well for the boundary conditions
considered. In particular, in the twisted sector a single-spinon
excitation to the conformal ground state arises which describes in
the continuum the topological defect occurring in a closed ladder
with MBC.

The role in determining different massive RG flows of the relevant
interactions$\ $($H_{\text{Railroad}}$, $H_{\text{4-Spin}}$) and
of
the marginal relevant interactions ($H_{\text{ZigZag}}$, $H_{\text{4-Dimer}}$%
) is investigated in the isotropic ladder case with $\Delta =1$.
In the four fermion description of the system, for both couples of
interactions, the result is the possibility to generate massive RG
flows with at will unbalanced fermion triplet ($m_{t}$) and
singlet ($m_{s}$) masses. The fact that, at the conformal point,
our TM model is the exact tensor product of
the degrees of freedom of the fermion triplet (i.e. the affine $\widehat{su(2)%
}_{2}$, $c=3/2$) and of the singlet (i.e. the $\bar{I}$ Ising
model, $c=1/2$ ) implies that along these RG flows the topological
structure of TM is preserved.\ In fact, the TM global symmetry
$su(2)\otimes Z_{2}$ cannot be broken by different values of the
triplet and singlet masses, while it can be enhanced to a
\textit{partial} conformal one $\widehat{su(2)}_{2}\otimes Z_{2}$
or $su(2)\otimes \bar{I}$ along the special RG flows with
$m_{t}=0$ or $m_{s}=0$, respectively. This stability allows us to
follow the evolution of the single-spinon excitation while it
acquires mass along a given RG flow. Thus, we can claim that the
existence of massive single-spinon excitations is a general
feature of closed ladders with MBC in the presence of various
types of interactions. In the strong massive limit then such a
single-spinon excitation should coincide with the one described in
Ref. \cite {mobius} in the special case of the zigzag interaction.

Finally, the possibility of getting a whole massless flow starting
from the UV fixed point with central charge $c=2$ is briefly
discussed. Further analysis on such an issue as well as the
complete RG analysis for the anisotropic ladder with $\Delta =0$
will be the subject of a forthcoming publication. Another issue
which deserves deeper investigation is the occurrence of a
topological order in such spin ladders of non trivial geometry.
That could result from the presence of a topological defect, in
close analogy with the Josephson junction ladders studied in Refs.
\cite{noi3,noi4,noi6}.

\section*{Acknowledgments}

We warmly thank Alan Luther for many enlightening discussions. G. N. is
supported by the contract MEXT-CT-2006-042695.

\section*{Appendix A: TM on the torus}

In this Appendix we summarize the primary field content of our TM model on
the torus topology for the particular case $m=2$ and for the
compactification radius $R_{\phi }^{2}=2$ of the mother theory, which
describes the 2-leg $XXX$ ladder. The decomposition of TM in terms of the $%
Z_{2}$-invariant (the affine $\widehat{su(2)}_{2}$ with $c=\frac{3}{2}$) and
$Z_{2}$-twisted (the Ising model with $c=\frac{1}{2}$ ) sub-theories is well
evidenced on the torus \cite{cgm4}\ \ by the corresponding decomposition of
the characters. In order to make it explicit let us start introducing the
characters of these two RCFT. We denote with $\bar{\chi}_{0}(\tau )$, $\bar{%
\chi}_{\frac{1}{2}}(\tau )$, $\bar{\chi}_{\frac{1}{16}}(\tau )$ the
characters of the chiral primary fields $I$, $\psi $ and $\sigma $ in the
Ising model \cite{cft} with Neveu-Schwartz ($Z_{2}$-twisted) boundary
conditions \cite{cgm4}, while
\begin{eqnarray}
\chi _{0}^{\widehat{su(2)}_{2}}(w|\tau ) &=&\chi _{0}(\tau )K_{0}(w|\tau
)+\chi _{\frac{1}{2}}(\tau )K_{2}(w|\tau )\,,  \label{mr1} \\
\chi _{1}^{\widehat{su(2)}_{2}}(w|\tau ) &=&\chi _{\frac{1}{16}}(\tau
)\left( K_{1}(w|\tau )+K_{3}(w|\tau )\right) ,  \label{mr2} \\
\chi _{2}^{\widehat{su(2)}_{2}}(w|\tau ) &=&\chi _{\frac{1}{2}}(\tau
)K_{0}(w|\tau )+\chi _{0}(\tau )K_{2}(w|\tau )  \label{mr3}
\end{eqnarray}
represent the characters of the three extended chiral primary fields in the
affine $Z_{2}$-invariant $\widehat{su(2)}_{2}$. The characters $\chi _{0,2}^{%
\widehat{su(2)}_{2}}$ contain only integer spin (i.e. two-spinon
excitations) while $\chi _{1}^{\widehat{su(2)}_{2}}$ describes half-integer
spin (i.e. single-spinon excitations). The above character formulae express
the decomposition of $\widehat{su(2)}_{2}$ in terms of the $c=1$ RCFT\ $%
\widehat{u(1)}_{4}$ (generated by the compactified boson $\mathcal{X}$ with $%
R_{\mathcal{X}}^{2}=1$) and in terms of the Ising model with Ramond boundary
conditions. In particular, the $K_{l}(w|\tau )$ defined by \cite{cgm4}:
\begin{equation}
K_{l}(w|\tau )=\frac{1}{\eta \left( \tau \right) }\;\Theta \left[
\begin{array}{c}
\frac{l}{4} \\[6pt]
0
\end{array}
\right] (2w|4\tau )\,,\qquad \text{with }l=0,1,2,3,  \label{chp}
\end{equation}
are the characters corresponding to the four extended chiral primary fields
of the RCFT\ $\widehat{u(1)}_{4}$ while we denote with $\chi _{i}(\tau )$
those of the Ising model with Ramond boundary conditions.

We can write now the characters\ of TM, splitting them in the four sectors
of the $Z_{2}$-orbifold. We have two extended chiral primaries in the $A-P$
sector with characters given by:
\begin{eqnarray}
\chi _{0}^{+}(w|\tau ) &=&\bar{\chi}_{\frac{1}{16}}(\tau )\left( \chi _{0}^{%
\widehat{su(2)}_{2}}(w|\tau )+\chi _{2}^{\widehat{su(2)}_{2}}(w|\tau
)\right) ,  \label{tw1} \\
\chi _{1}^{+}(w|\tau ) &=&\left( \bar{\chi}_{0}(\tau )+\bar{\chi}_{\frac{1}{2%
}}(\tau )\right) \chi _{1}^{\widehat{su(2)}_{2}}(w|\tau ),  \label{tw2}
\end{eqnarray}
and two extended chiral primaries in the $A-A$ sector whose characters read:
\begin{eqnarray}
\chi _{0}^{-}(w|\tau ) &=&\bar{\chi}_{\frac{1}{16}}(\tau )\left( \chi _{0}^{%
\widehat{su(2)}_{2}}(w|\tau )-\chi _{2}^{\widehat{su(2)}_{2}}(w|\tau
)\right) ,  \label{tw3} \\
\chi _{1}^{-}(w|\tau ) &=&\left( \bar{\chi}_{0}-\bar{\chi}_{\frac{1}{2}%
}\right) (\tau )\chi _{1}^{\widehat{su(2)}_{2}}(w|\tau ).  \label{tw4}
\end{eqnarray}
In the $P-A$ sector there are two extended chiral primaries with the
following characters:
\begin{align}
\tilde{\chi}_{0}^{-}(w|\tau )& =\bar{\chi}_{0}(\tau )\chi _{0}^{\widehat{%
su(2)}_{2}}(w|\tau )-\bar{\chi}_{\frac{1}{2}}(\tau )\chi _{2}^{\widehat{su(2)%
}_{2}}(w|\tau ),  \label{vac1} \\
\tilde{\chi}_{1}^{-}(w|\tau )& =\bar{\chi}_{0}(\tau )\chi _{2}^{\widehat{%
su(2)}_{2}}(w|\tau )-\bar{\chi}_{\frac{1}{2}}(\tau )\chi _{0}^{\widehat{su(2)%
}_{2}}(w|\tau ),  \label{vac2}
\end{align}
while for the $P-P$ sector we have three extended chiral primaries with
characters given by:
\begin{align}
\tilde{\chi}_{\alpha }^{+}(w|\tau )& =\bar{\chi}_{0}(\tau )\chi _{0}^{%
\widehat{su(2)}_{2}}(w|\tau )+\bar{\chi}_{\frac{1}{2}}(\tau )\chi _{2}^{%
\widehat{su(2)}_{2}}(w|\tau )\,,  \label{vac3} \\
\tilde{\chi}_{\beta }^{+}(w|\tau )& =\bar{\chi}_{0}(\tau )\chi _{2}^{%
\widehat{su(2)}_{2}}(w|\tau )+\bar{\chi}_{\frac{1}{2}}(\tau )\chi _{0}^{%
\widehat{su(2)}_{2}}(w|\tau ),  \label{vac4} \\
\tilde{\chi}_{\gamma }^{+}(w|\tau )& =\bar{\chi}_{\frac{1}{16}}(\tau )\chi
_{1}^{\widehat{su(2)}_{2}}(w|\tau );  \label{vac5}
\end{align}
note that the above factorization expresses the parity selection rule ($m$%
-ality). It is interesting to point out that in the $P-P$ sector, unlike for
the other sectors, modular invariance constraint requires the presence of
three different characters. The isospin operator content of the character $%
\tilde{\chi}_{\gamma }^{+}(w|\tau )$ clearly evidences its peculiarity with
respect to the other states of the periodic case. Indeed it is characterized
by two twist fields ($\Delta =1/16$) in the isospin components. The
occurrence of the double twist in such a state is simply due to periodicity.
Indeed, being an isospin twist field the representation in the continuum
limit of a single-spinon excitation, the double twist corresponds to a
two-spinon excitation, typical of the periodic configuration.

Finally, it is possible to show (see \cite{cgm4})\ that the
diagonal partition function of the TM on the torus has the
following factorized form:
\begin{equation}
Z(w|\tau )=Z^{\widehat{su(2)}_{2}}(w|\tau )Z_{\overline{I}}(\tau )\text{,}
\label{partition1}
\end{equation}
in terms of the $\widehat{su(2)}_{2}$\ partition function $Z^{\widehat{su(2)}%
_{2}}$ ($c=3/2$)\ and of the Ising partition function $Z_{\overline{I}}$ ($%
c=1/2$). They have, in turn, the following expression:
\begin{eqnarray}
Z^{\widehat{su(2)}_{2}}(w|\tau ) &=&|\chi _{0}^{\widehat{su(2)}_{2}}(w|\tau
)|^{2}+|\chi _{1}^{\widehat{su(2)}_{2}}(w|\tau )|^{2}+|\chi _{2}^{\widehat{%
su(2)}_{2}}(w|\tau )|^{2}, \\
&&  \notag \\
Z_{\overline{I}}(\tau ) &=&|\bar{\chi}_{0}(\tau )|^{2}+|\bar{\chi}_{\frac{1}{%
2}}(\tau )|^{2}+|\bar{\chi}_{\frac{1}{16}}(\tau )|^{2}.
\end{eqnarray}

\section*{Appendix B: mass analysis within the self-consistent mean field
approximation}

In this Appendix we carry out an analysis in a self-consistent mean field
approximation\ for the masses $m_{t}$ and $m_{s}$. Our situation is more
complex than the one quoted in Ref. \cite{allen} because we deal with the
case where either $\lambda _{+}$ or $\lambda _{-}$ could be marginally
relevant and provide a dynamical contribution to the singlet and triplet
masses. So we need to keep track of a new feature: the spin singlet and
triplet are in the presence of different interactions. Then we may write:
\begin{equation}
\psi _{0}(z)\bar{\psi}_{0}(\bar{z})\rightarrow i\varepsilon _{s}+\psi _{0}(z)%
\bar{\psi}_{0}(\bar{z}),\text{ \ \ }\psi _{j}(z)\bar{\psi}_{j}(\bar{z}%
)\rightarrow i\varepsilon _{t}+\psi _{j}(z)\bar{\psi}_{j}(\bar{z}),\text{ \ }%
j=1,2,3,  \label{per30}
\end{equation}
where $\varepsilon _{s}$ and $\varepsilon _{t}$ can be a priori different.
By substituting the expressions (\ref{per30}) in the general perturbing term
$\mathcal{V}_{\text{Tot}}$, Eq. (\ref{per28}), and neglecting terms quartic
in the fields $\psi $, we get:
\begin{equation}
\mathcal{V}_{\text{Tot}}=i\sum\limits_{i=1}^{3}m_{t}\psi _{i}(z)\bar{\psi}%
_{i}(\bar{z})+im_{s}\psi _{0}(z)\bar{\psi}_{0}(\bar{z})  \label{per31}
\end{equation}
where:
\begin{equation}
m_{t}=2\pi (2\varepsilon _{t}\Lambda _{1}+\varepsilon _{s}\Lambda _{2}),%
\text{ \ \ }m_{s}=6\pi \varepsilon _{t}\Lambda _{2}.  \label{per32}
\end{equation}
$\varepsilon _{t}$ and $\varepsilon _{s}$ may be determined
self-consistently by using the expressions for the correlation functions of
massive fermion fields; we get:
\begin{equation}
\varepsilon _{s}=m_{s}\int \frac{d^{2}k}{\pi }\frac{e^{-i\text{k}\cdot \text{%
x}}}{\text{k}^{2}+m_{s}^{2}}\text{ \ \ , \ }\varepsilon _{t}=m_{t}\int \frac{%
d^{2}k}{\pi }\frac{e^{-i\text{k}\cdot \text{x}}}{\text{k}^{2}+m_{t}^{2}}
\label{per33}
\end{equation}
with $m_{t}$ and $m_{s}$ replaced by the expressions given in Eqs. (\ref
{per32}).

Along the special RG trajectory characterized by the condition $\Lambda
_{2}=0$, the mass $m_{s}$ vanishes and the above system of equations reduces
to the single equation in the triplet mass $m_{t}$:
\begin{equation}
\frac{m_{t}}{4\pi \Lambda _{1}}=m_{t}\int \frac{d^{2}k}{\pi }\frac{e^{-i%
\text{k}\cdot \text{x}}}{\text{k}^{2}+m_{t}^{2}}\text{ }
\end{equation}
whose non-zero solution is given by (\ref{per29}).


\begin{thebibliography}{99}
\bibitem{cgm4}  G. Cristofano, G. Maiella, V. Marotta, \textit{Mod. Phys.
Lett. A }\textbf{15} (2000) 547; G. Cristofano, G. Maiella, V. Marotta,
\textit{Mod. Phys. Lett. A}\textbf{\ 15} (2000) 1679; G. Cristofano, G.
Maiella, V. Marotta, G. Niccoli, \textit{Nucl. Phys. B }\textbf{641 }(2002)
547; G. Cristofano, V. Marotta, G. Niccoli, \textit{JHEP }\textbf{06 }(2004)
056.

\bibitem{haldane}  F. D. M. Haldane, \textit{Phys. Rev. Lett. }\textbf{50}
(1983) 1153; F. D. M. Haldane, \textit{Phys. Lett. A }\textbf{93} (1983) 464.

\bibitem{ladder}  D. C. Johnston, J. W. Johnson, D. P. Goshorn, A. J.
Jacobson, \textit{Phys. Rev. B }\textbf{35 }(1987) 219; Z. Hiroi, M. Azuma,
M. Takano, Y. Bando, \textit{J. Solid State Chem. }\textbf{95} (1991) 230;
E. Dagotto, \textit{Rep. Prog. Phys. }\textbf{62 }(1999) 1525.

\bibitem{liquid}  E. Dagotto, T. M. Rice, \textit{Science }\textbf{271}
(1996) 618; Y. Wang, \textit{Phys. Rev. B }\textbf{60 }(1999) 9236.

\bibitem{plateaux}  K. Hida, \textit{J. Phys. Soc. Jpn. }\textbf{63} (1994)
2359; T. Tonegawa, T. Nakao, M. Kaburagi, \textit{J. Phys. Soc. Jpn. }%
\textbf{65} (1996) 3317; M. Oshikawa, M. Yamanaka, I. Affleck, \textit{Phys.
Rev. Lett. }\textbf{78} (1997) 1984; D. C. Cabra, A. Honecker, P. Pujol,
\textit{Eur. Phys. J. B }\textbf{13 }(2000) 55.

\bibitem{csl}  N. Andrei, M. R. Douglas, A. Jerez, \textit{Phys. Rev. B }%
\textbf{58 }(1998) 7619; P. Azaria, P. Lecheminant, \textit{Nucl. Phys. B }%
\textbf{575 }(2000) 439; P. Azaria, P. Lecheminant, A. A. Nersesyan, \textit{%
Phys. Rev. B }\textbf{58 }(1998) R8881.

\bibitem{railroad}  S. P. Strong, A. J. Millis, \textit{Phys. Rev. Lett. }%
\textbf{69} (1992) 2419; S. P. Strong, A. J. Millis, \textit{Phys. Rev. B }%
\textbf{50} (1994) 9911; M. Sigrist, T. M. Rice, F. C. Zhang, \textit{Phys.
Rev. B }\textbf{49 }(1994) 12058; S. R. White, R. M. Noack, D. J. Scalapino,
\textit{Phys. Rev. Lett. }\textbf{73} (1994) 886.

\bibitem{scalapino}  D. J. Scalapino, \textit{Nature }\textbf{377} (1995)
12; E. Dagotto, J. Riera, D. J. Scalapino, \textit{Phys. Rev. B }\textbf{45 }%
(1992) 5744.

\bibitem{shelton1}  D. G. Shelton, A. A. Nersesyan, A. M. Tsvelik, \textit{%
Phys. Rev. B }\textbf{53 }(1996) 8521; D. C. Cabra, A. Dobry, G. L. Rossini,
\textit{Phys. Rev. B }\textbf{63 }(2001) 144408.

\bibitem{wang1}  A. A. Nersesyan, A. M. Tsvelik, \textit{Phys. Rev. Lett. }%
\textbf{78 }(1997) 3939; Y. J. Wang, \textit{Phys. Rev. B }\textbf{68 }%
(2003) 214428.

\bibitem{haldane1}  F. D. M. Haldane, \textit{Phys. Rev. B }\textbf{25}
(1982) R4925; F. D. M. Haldane, \textit{Phys. Rev. B }\textbf{26} (1982)
5257.

\bibitem{shastry}  B. S. Shastry, B. Sutherland, \textit{Phys. Rev. Lett. }%
\textbf{47} (1981) 964.

\bibitem{exper}  M. Azuma, Z. Hiroi, M. Takano, K. Ishida, Y. Kitaoka,
\textit{Phys. Rev. Lett. }\textbf{73} (1994) 3463; G. Castilla, S.
Chakravarty, V. J. Emery, \textit{Phys. Rev. Lett. }\textbf{75} (1995) 1823.

\bibitem{exper1}  W. Shiramura, K. Takatsu, H. Tanaka, K. Kamishima, M.
Takahashi, H. Mitamura, T. Goto, \textit{J. Phys. Soc. Jpn. }\textbf{66}
(1997) 1900.

\bibitem{exper2}  W. Shiramura, K. Takatsu, B. Kurniawan, H. Tanaka, H.
Uekusa, Y. Ohashi, K. Takizawa, H. Mitamura, T. Goto, \textit{J. Phys. Soc.
Jpn. }\textbf{67} (1998) 1548.

\bibitem{chiral}  A. A. Nersesyan, A. O. Gogolin, F. H. L. Essler, \textit{%
Phys. Rev. Lett. }\textbf{81} (1998) 910; P. Lecheminant, T. Jolicoeur, P.
Azaria, \textit{Phys. Rev. B }\textbf{63 }(2001) 174426; T. Hikihara, M.
Kaburagi, H. Kawamura, \textit{Phys. Rev. B }\textbf{63 }(2001) 174430.

\bibitem{gjja}  E. Altman, A.\ Auerbach, \textit{Phys. Rev. Lett. }\textbf{81%
} (1998) 4484.

\bibitem{white1}  S. R. White, I. Affleck, \textit{Phys. Rev. B }\textbf{54 }%
(1996) 9862.

\bibitem{moxides}  S. Yunoki, J. Hu, A. L. Malvezzi, A. Moreo, N. Furukawa,
E. Dagotto, \textit{Phys. Rev. Lett. }\textbf{80} (1998) 845.

\bibitem{mobius}  K. Okunishi, N. Maeshima, \textit{Phys. Rev. B }\textbf{64
}(2001) 212406.

\bibitem{grimm}  U. Grimm, \textit{J. Phys. A: Math. Gen. }\textbf{35}
(2002) L25.

\bibitem{topological}  X. G. Wen, F. Wilczek, A. Zee, \textit{Phys. Rev. B }%
\textbf{39 }(1989) 11413; X. G. Wen, \textit{Phys. Rev. B }\textbf{44 }%
(1991) 2664; T. Senthil, M. P. A. Fisher, \textit{Phys. Rev. B }\textbf{61 }%
(2000) 9690; E. H. Kim, O. Legeza, J. Solyom, \textit{Phys. Rev. B }\textbf{%
77 }(2008) 205121.

\bibitem{wen}  X. G. Wen, \textit{Int. J. Mod. Phys.\ B} \textbf{6 } (1992)
1711; X. G. Wen, \textit{Adv. in Phys.} \textbf{44} (1995) 405.

\bibitem{noi1}  G. Cristofano, V. Marotta, A. Naddeo, \textit{Phys. Lett. B }%
\textbf{571} (2003) 250.

\bibitem{noi2}  G. Cristofano, V. Marotta, A. Naddeo, \textit{Nucl. Phys. B }%
\textbf{679 }(2004) 621.

\bibitem{noi5}  G. Cristofano, V. Marotta, A. Naddeo, G. Niccoli, \textit{J.
Stat. Mech.: Theor. Exper. }(2006) L05002.

\bibitem{noi3}  G. Cristofano, V. Marotta, A. Naddeo, \textit{J. Stat.
Mech.: Theor. Exper. }(2005) P03006.

\bibitem{noi4}  G. Cristofano, V. Marotta, A. Naddeo, G. Niccoli, \textit{%
Eur. Phys. J. B}\textbf{\ 49 }(2006) 83.

\bibitem{noi6}  G. Cristofano, V. Marotta, A. Naddeo, G. Niccoli, \textit{%
Phys. Lett. A }\textbf{372} (2008) 2464.

\bibitem{noi}  G. Cristofano, V. Marotta, P. Minnhagen, A. Naddeo, G.
Niccoli, \textit{J. Stat. Mech.: Theor. Exper. }(2006) P11009.

\bibitem{bos}  I. Affleck, in \textit{Fields, Strings and Critical Phenomena}
(Les Houches, Session XLIX), E. Br\'{e}zin and J. Zinn-Justin (Eds.),
Amsterdam: North-Holland (1988); A. O. Gogolin, A. A. Nersesyan, A. M.
Tsvelick, \textit{Bosonization and Strongly Correlated Systems} (Cambridge
University Press, Cambridge, 1998); H. J. Schulz, \textit{Phys. Rev. B }%
\textbf{34 }(1986) 6372.

\bibitem{betheansatz}  J. D. Johnson, S. Krinsky, B. M. McCoy, \textit{Phys.
Rev. A }\textbf{8 }(1973) 2526; A. Luther, I. Peschel, \textit{Phys. Rev.\ B}
\textbf{12} (1975) 3908.

\bibitem{VM}  V. Marotta, \textit{J. Phys. A\ }\textbf{26} (1993) 3481; V.
Marotta, \textit{\ Mod. Phys. Lett.\ A}\textbf{\ 13} (1998) 853; V. Marotta,
\textit{\ Nucl. Phys.\ B }\textbf{527} (1998) 717; V. Marotta, \textit{Mod.
Phys. Lett.\ A}\textbf{13} (1998) 2863.

\bibitem{noi7}  G. Cristofano et al., work in preparation.

\bibitem{nomura}  K. Hijii, A. Kitazawa, K. Nomura, \textit{Phys. Rev.\ B}
\textbf{72} (2005) 014449.

\bibitem{allen}  D. Allen, D. Senechal, \textit{Phys. Rev. B }\textbf{55 }%
(1997) 299.

\bibitem{itoi}  C. Itoi, S. Qin, \textit{Phys. Rev. B }\textbf{63 }(2001)
224423.

\bibitem{zigcritical}  K. Okamoto, K. Nomura, \textit{Phys. Lett.\ A}
\textbf{169} (1993) 433; S. Eggert, \textit{Phys. Rev.\ B} \textbf{54}
(1996) R9612.

\bibitem{ghosh}  C. K. Majumdar, D. K. Ghosh, \textit{J. Math. Phys.}
\textbf{10} (1969) 1388; C. K. Majumdar, D. K. Ghosh, \textit{J. Math. Phys.}
\textbf{10} (1969) 1399.

\bibitem{kolezuk}  A. K. Kolezhuk, H. J. Mikeska, \textit{Phys. Rev.\ B}
\textbf{56} (1997) R11380.

\bibitem{incommensurate}  T. Tonegawa, I. Harada, M. Kaburagi, \textit{J.
Phys. Soc. Jpn. }\textbf{61} (1992) 4665; R. Bursill, G. A. Gehring, D. J.
J. Farnell, J. B. Parkinson, Tao Xiang, Chen Zeng, \textit{J. Phys.:
Condens. Matt.} \textbf{7} (1995) 8605; A. A. Aligia, C. D. Batista, F. H.
L. Essler, \textit{Phys. Rev.\ B} \textbf{62} (2000) 3259; O. Legeza, J.
Solyom, L. Tincani, R. M. Noack, \textit{Phys. Rev.\ Lett.} \textbf{99}
(2007) 087203.

\bibitem{sorensen}  E. Sorensen, I. Affleck, D. Augier, D. Poilblanc,
\textit{Phys. Rev. B }\textbf{58 }(1998) R14701.

\bibitem{Giam}  T. Giamarchi, \textit{Quantum Physics in One Dimension},
Clarendon Press, Oxford (2004).

\bibitem{A-M}  R. G. Pereira, J. Sirker, J. S. Caux, R. Hagemans, J. M.
Maillet, S. R. White and I. Affleck, \textit{J. Stat. Mech.: Theor. Exper. }%
(2007) P08022.

\bibitem{L-T}  S. Lukyanov and V. Terras, \textit{Nucl. Phys.} \textit{B}
\textbf{654} (2003) 323.

\bibitem{nishi}  Y. Nishiyama, \textit{Eur. Phys. J. B }\textbf{17} (2000)
295.

\bibitem{nbos}  I. Affleck, \textit{Nucl. Phys. B }\textbf{265 }(1986) 409;
E. Witten, \textit{Comm. Math. Phys. }\textbf{92 }(1984) 455; A. B.
Zamolodchikov, V. A. Fateev, \textit{Sov. J. Nucl. Phys. }\textbf{43 }(1986)
657; V. G. Knizhnik, A. B. Zamolodchikov, \textit{Nucl. Phys. B }\textbf{247
}(1984) 83.

\bibitem{cft}  P. Di Francesco, P. Mathieu, D. Senechal, \textit{Conformal
Field Theories}, Springer-Verlag, (1996).
\end{thebibliography}
\end{document}